\documentclass[11pt]{article}
\usepackage[T1]{fontenc}
\usepackage{lmodern,amsmath,amssymb,amsthm,mathtools,booktabs}
\usepackage[margin=1in]{geometry}
\usepackage{placeins}
\usepackage[round,authoryear]{natbib}
\usepackage{graphicx,xcolor}
\usepackage[colorlinks=true,allcolors=blue!45!black]{hyperref}
\hypersetup{pdftitle={Marked INAR order flow and the rough Hawkes-Heston limit},
 pdfsubject={Joint convergence of marked INAR order flow to the rough Hawkes-Heston model}}
\numberwithin{equation}{section}
\newtheorem{theorem}{Theorem}[section]
\newtheorem{proposition}[theorem]{Proposition}
\newtheorem{lemma}[theorem]{Lemma}
\newtheorem{corollary}[theorem]{Corollary}
\newcommand{\E}{\mathbb E}
\newcommand{\Prob}{\mathbb P}
\newcommand{\R}{\mathbb R}

\newcommand{\F}{\mathcal F}
\newcommand{\dd}{\mathrm d}
\newcommand{\ii}{\mathrm i}
\newcommand{\J}{\mathcal J}
\newcommand{\e}{\mathrm e}
\newcommand{\ind}{\mathbf1}
\newcommand{\K}{K_\alpha}
\newcommand{\G}{G_\alpha}
\newcommand{\sig}{\{+,-\}}
\newcommand{\spc}{(0,\infty)}

\newcommand{\Law}{\mathcal L}
\allowdisplaybreaks[2]

\title{Marked INAR model and the rough Hawkes--Heston limit%
\thanks{The code used in the numerical experiments is available on
\href{https://github.com/gagawjbytw/marked-inar-rough-hawkes-heston}{GitHub}.}}

\author{
Xiaoyu Wang\thanks{FinTech Thrust, Hong Kong University of Science and Technology (Guangzhou), Guangzhou, Guangdong, People's Republic of China; \texttt{xiaoyuwang@hkust-gz.edu.cn}}
\and
Yingli Wang\thanks{Corresponding Author. School of Mathematical Sciences, Fudan University, Shanghai, People's Republic of China.\newline\texttt{yingliwang@fudan.edu.cn}}
\and
Lingjiong Zhu\thanks{Department of Mathematics, Florida State University, Tallahassee, Florida, United States of America.\newline\texttt{zhu@math.fsu.edu}}}

\date{\today}

\begin{document}
\maketitle

\begin{abstract}
We develop a discrete-time Monte Carlo algorithm for the rough Hawkes--Heston
model based on a nearly unstable marked integer-valued autoregressive (INAR)
process. We implement the algorithm with a block fast Fourier transform (FFT)
scheme that evaluates the excitation convolution in $\mathcal O(N\log^2 N)$
operations per path on an $N$-bin grid.
We prove joint convergence in law of the spot variance, log price and common
marked point measure to their rough Hawkes--Heston limits.
We prove convergence of European call and put option prices, as well as convergence
of arithmetic Asian and down-and-out put option prices to their continuously
monitored values.
For European options, the one-year implied volatility smiles computed from
the grid transform approach the continuous affine reference across four
roughness levels. Monte Carlo experiments on the same three grids compare
arithmetic Asian and down-and-out put option prices with a finer-grid reference
that demonstrate the efficiency of the proposed algorithm.
\end{abstract}

\section{Introduction}\label{sec:intro}

Hawkes processes are self-exciting point processes whose conditional event
intensity is the sum of a baseline rate and nonnegative kernel contributions
from past events \citep{hawkes}. In multivariate models, events of one type
can also increase the arrival rates of other types, allowing the model to
capture both clustering and interactions between event streams.
They are widely used in finance to model trades, price changes and order-book
dynamics \citep{hawkes-finance}. For example, \citet{bacry-microstructure}
use mutually exciting point processes to model microstructure noise and
dependence between asset prices, while \citet{morariu-pakkanen} use
state-dependent Hawkes processes to describe the feedback between order flow
and the state of the limit order book.

An order that changes a quote can also alter the rate of subsequent orders.
Marked Hawkes processes describe this feedback by attaching a size or type to each
event. Let $\mathcal N(\dd s,\dd z)$ record event times and marks in a space $E$,
and let $F$ be the law of a new mark, sampled independently of the past.
For a nonnegative baseline $\mu$, lag kernel $\phi$ and mark response $g:E\rightarrow\mathbb{R}_{+}$,
a separable marked Hawkes model has the predictable intensity:
\[
 \lambda_t=\mu(t)
   +\int_{[0,t)\times E}\phi(t-s)g(z)\,\mathcal N(\dd s,\dd z),
\]
and marked compensator $\lambda_t\dd t\,F(\dd z)$.
The choice $g\equiv1$ recovers the linear Hawkes model of \citet{hawkes}.
In the immigration-birth (branching) interpretation of \citet{clusters}, a parent with mark $z$
produces offspring at rate $\phi(u)g(z)$ at lag $u>0$. Thus the same mark
can determine both an immediate price response and the excitation of future orders.
\citet{KarabashZhu} study the limit theorems for marked Hawkes processes.
\citet{hxmarked} establish functional laws of large numbers and central limit
theorems for marked Hawkes point measures and their associated shot-noise processes.

There is a discrete-time analogue of the marked Hawkes model. 
For observations recorded in bins, the primitive quantities are integer counts.
Integer-valued autoregressions (INAR) propagate counts by thinning \citep{inar}.
A marked INAR($\infty$) model with Poisson offspring assigns a mark
$\zeta_{j,k}$ to the $k$th event in bin $j$. Given the history
$\mathcal H_{n-1}$ of counts and marks through bin $n-1$, it satisfies
\[
 N_n\mid\mathcal H_{n-1}\sim\operatorname{Pois}\!\left(
   \mu_n+\sum_{j=0}^{n-1}\phi_{n-j}
               \sum_{k=1}^{N_j}g(\zeta_{j,k})\right),\qquad n\ge1.
\]
Similar to the Hawkes process \citep{clusters}, 
there is also an immigration-birth representation for the INAR model. 
Here immigrants in bin $n$ have Poisson mean $\mu_n\geq 0$, and each parent
contributes an independent Poisson number of offspring with mean
$\phi_{n-j}g(\zeta_{j,k})$, where $\phi_{n-j}>0$ and $g:E\rightarrow\mathbb{R}_{+}$. New marks are sampled independently from $F$.
The counts remain integer valued even when the marks and their excitation weights
are real valued. For $g\equiv1$, this is the Poisson-thinning INAR recursion
used by \citet{kirchner} to approximate the continuous-time Hawkes processes.
With mark-dependent excitation, it is a discrete-time marked Hawkes model of the
type studied by \citet{wangmarked}, who proves laws of large numbers and central
limit theorems. The model below combines signed regular counts and rare marked
counts, with their frequencies and excitation weights depending on the mesh.

Nearly unstable self-excitation also provides a mechanism for stochastic volatility.
\citet{jr15} obtain a square-root diffusion from nearly unstable Hawkes processes,
and \citet{jr16} obtain fractional dynamics from heavy-tailed excitation.
\citet{rough} document the roughness of log volatility. \citet{leverage} connect
asymmetric order flow to leverage and rough volatility. In a Hawkes order-flow
framework, \citet{jrimpact} relate the no-arbitrage condition to power-law market
impact and rough volatility at the macroscopic scale.
\citet{hxzrough} prove convergence of the rescaled heavy-tailed Hawkes intensity
itself and joint convergence of price and volatility to the rough Heston model, using
$C$-tightness estimates for the intensity processes.

Let $S_t$ be the discounted stock price with $S_0>0$, $X_t:=\log(S_t/S_0)$
the log return, and $A_t:=\int_0^t V_s\dd s$ the quadratic variation of its
continuous martingale part, where $V_t$ is the variance process. We call $V$ the spot variance. The rough Heston model
of \citet{rh} uses the fractional kernel $\K(t):=t^{\alpha-1}/\Gamma(\alpha)$
for $t>0$ and $1/2<\alpha<1$, whose dynamics is given by:
\begin{align}
 X_t&=\int_0^t\sqrt{V_s}\dd B^S_s-\frac12 A_t,\label{rough:Heston:1}\\
 V_t&=G_0(t)+b\int_0^t\K(t-s)V_s\dd s
       +\sqrt c\int_0^t\K(t-s)\sqrt{V_s}\dd B^V_s,\label{rough:Heston:2}
\end{align}
where the initial curve $G_0$ is deterministic, continuous and nondecreasing
on $[0,\infty)$, with $G_0(0)\ge0$. The marked INAR scaling below takes
$G_0(t)=v_0+\eta t^\alpha/\Gamma(1+\alpha)$, with $v_0,\eta\ge0$.
Here $b<0$ is the mean-reversion
coefficient, and $\sqrt c$ is the volatility of volatility. 
$B^S$ and $B^V$ are standard Brownian motions 
that satisfy $\dd\langle B^S,B^V\rangle_t=\rho\,\dd t$. A negative
$\rho$ gives leverage, and $\alpha-\frac12$ governs the roughness of $V$.

In discrete time, \citet{cai} derive a fractional diffusion limit from a univariate
cumulative INAR process. \citet{wcz} obtain the joint rough Heston price and variance
from bivariate counts and derive an exact lattice transform for European option pricing.
Their Section~4.3 compares grid-monitored Asian and barrier option prices with a
classical Heston benchmark at $\alpha=1$. In the diffusion-driven
Volterra--Heston model, \citet{custers-asian} study discretely monitored
arithmetic and geometric Asian options.

For macroscopic jump dynamics, \citet{hxjumps} derive stochastic volatility models
with self-exciting price--volatility co-jumps from microscopic order flow, with
separate Poisson random measures for exogenous and endogenous shocks.
\citet{hxzspikes} combine persistent market-order impact and temporary
limit-order impact to obtain rough volatility driven by an additional Poisson
random measure, producing volatility spikes and clusters of spikes.

\citet{rh} express the rough Heston characteristic function through a fractional
Riccati equation. \citet{affinevolterra} derive Fourier--Laplace functionals for
affine Volterra processes. \citet{abijaberL1} establishes weak existence,
uniqueness and stability for affine stochastic Volterra equations with locally
integrable kernels and jumps. In the rough Hawkes--Heston model,
\citet[Sections~3--4]{bps} obtain transform formulas for European stock options
and VIX derivatives.
A down-and-out option also records whether a price decline
at an intermediate time crossed its barrier. For example, a common mark of size
$z$ knocks the stock below a level $L$ whenever
$S_{t-}>L>S_{t-}\e^{-\Lambda z}$, regardless of the price at maturity.

\citet{wwz} obtain this model from a continuous-time heavy-tailed
marked Hawkes system. Their Proposition~4.8 proves $C$-tightness of the centered
regular order flow. Their Theorem~5.1 gives joint convergence of the spot variance
in $L^2(0,1)$, the log price in $J_1$ and the common jump measure in the vague
topology. Their Figure~2 compares two short-dated SPX smiles.

The rough Hawkes--Heston model of \citet{bps} adds a common marked point measure
$\J$ to the two rough Heston equations \eqref{rough:Heston:1}-\eqref{rough:Heston:2}. Its predictable compensator is
$V_s\dd s\,\nu(\dd z)$. We consider a finite measure $\nu$ on $\spc$
with finite second moment and write $\widetilde{\J}(\dd s,\dd z)
:=\J(\dd s,\dd z)-V_s\dd s\,\nu(\dd z)$. In compensated form the equations are
\begin{equation}\label{eqn:rough:Hawkes:Heston}
\begin{aligned}
 X_t&=\int_0^t\sqrt{V_s}\dd B^S_s
      -\left[\frac12+\int_{\spc}\left(\e^{-\Lambda z}-1+\Lambda z\right)\nu(\dd z)\right]A_t-\Lambda\int_{(0,t]\times\spc}z\,\widetilde{\J}(\dd s,\dd z),\\
 V_t&=G_0(t)+b\int_0^t\K(t-s)V_s\dd s
       +\sqrt c\int_0^t\K(t-s)\sqrt{V_s}\dd B^V_s
       \\
       &\qquad\qquad\qquad\qquad\qquad\qquad\qquad+\xi\int_{[0,t)\times\spc}\K(t-s)z\,\widetilde{\J}(\dd s,\dd z),
\end{aligned}
\end{equation}
where $B^S,B^V$ are standard Brownian motions with correlation $\rho$, 
$G_0(0)=v_0$ and $G_0(t)-G_0(s)=\eta\int_s^t\K(u)\dd u\ge0$ for $0\le s<t$, 
which satisfies \citet[Hypothesis~2.2]{bps},
$b<0$ is the mean-reversion
coefficient, and $\sqrt c$ is the volatility of volatility. 
A mark $z$ at time $s$ changes the log price by $-\Lambda z$, with
$\Lambda\ge0$. Its contribution to the raw mark integral in the variance
equation is the function $t\mapsto\xi z\K(t-s)$ for $t>s$, with $\xi>0$.
This term enters the spot variance that governs the arrival rate of future marks.
The variance equation holds for almost every $t$. Setting $\nu=0$ recovers the rough
Heston model \eqref{rough:Heston:1}-\eqref{rough:Heston:2}. 
Equations~\eqref{eq:target}--\eqref{eq:tprice} express the jump model
in terms of its martingale drivers and raw marks.

However, the rough Hawkes--Heston model \eqref{eqn:rough:Hawkes:Heston}
is a continuous-time model, and in practice one needs an algorithm to implement it.
Recently, \citet[Chapter~3]{serva} develops a Monte Carlo scheme for the rough
Hawkes--Heston model using a multifactor approximation of the fractional
kernel, Hawkes jump simulation and Euler time stepping for the log price.
In our paper, we adopt a novel approach by proposing a marked INAR process
as a practically implementable algorithm that can approximate
the continuous-time rough Hawkes--Heston model \eqref{eqn:rough:Hawkes:Heston}.
We establish rigorous convergence results to provide
a solid theoretical foundation. 
Our numerical scheme starts from a marked INAR process that generates counts
and marks bin by bin, with events in the same bin sharing a recorded time.
To our knowledge, this is the first numerical scheme based on a marked INAR model for the
rough Hawkes--Heston model with a rigorous proof of joint convergence of the spot
variance, log price and common marked point measure. We implement the scheme
with a divide-and-conquer FFT algorithm for the excitation convolution
(Section~\ref{subsec:path-numerics}).

\paragraph{Contributions}
\begin{enumerate}
\item We propose a novel discrete-time Monte Carlo algorithm for the rough Hawkes--Heston
model based on a nearly unstable marked integer-valued autoregressive (INAR)
process (Section~\ref{sec:model}). Theorem~\ref{thm:limit} proves that the spot variance, log price and common
marked point measure of the marked INAR model converge jointly in law to their
rough Hawkes--Heston limits. The respective topologies are the norm topology
of $L^2(0,T)$, the Skorokhod $J_1$ topology and the vague topology.
A limiting mark of size $z$ at time $s$ produces a log-price jump
$-\Lambda z$ and the response $\xi z\K(t-s)$, $t>s$, in the variance convolution.
\item Corollary~\ref{cor:european-prices} proves convergence of European call
and put option prices. We compute one-year European option implied volatility smiles from the marked
INAR transform and compare them with the continuous affine reference
(Figure~\ref{fig:smiles}). At each of the four roughness levels, the maximum
absolute difference in implied volatility over the eleven strikes decreases
as the grid is refined from $128$ to $512$ to $2{,}048$ bins
(Table~\ref{tab:grid}).
\item For the path-dependent options covered by
Corollary~\ref{cor:path-prices}, the option prices computed on the marked INAR grid
converge to the continuously monitored prices as the mesh tends to zero.
To the best of our knowledge, this is the first convergence result for path-dependent
option prices in the rough Hawkes--Heston model with marks that drive both
price jumps and variance excitation.
For the down-and-out put option, the convergence holds at every lower barrier
$L\in(0,S_0)$ because the limiting running minimum has no atom there
(Lemma~\ref{lem:barrier-atom}).
For the arithmetic Asian put options and down-and-out put options in Table~\ref{tab:path-refinement},
the $2{,}048$-bin Monte Carlo estimates differ from their $16{,}384$-bin
references by $0.4$--$2.4$ basis points of spot in absolute value.
\end{enumerate}

\paragraph{Organization of the paper}
Section~\ref{sec:main} constructs the marked order flow model using a marked INAR process and states its joint scaling
limit. Section~\ref{sec:European:prices} proves convergence
of European option prices and Section~\ref{sec:path-prices} proves convergence of path-dependent option prices.
Section~\ref{subsec:smiles} compares the one-year European option implied volatility
smiles of the grid and limiting models. Section~\ref{subsec:path-numerics}
examines Asian and down-and-out put option prices across monitoring grids.
Section~\ref{sec:end} concludes. 
All the technical proofs are presented in Appendix~\ref{app:proofs}.

\paragraph{Notation}
Fix a horizon $T>0$, let $h=T/N$, $N\in\mathbb N$, and let $t_n=nh$.
For one-sided sequences and functions, respectively, write
\[
 (f*g)_n:=\sum_{j=0}^n f_{n-j}g_j,
 \qquad (f*g)(t):=\int_0^t f(t-s)g(s)\dd s.
\]
The integrated fractional kernel is
\[
 \G(t):=\int_0^t\K(s)\dd s=\frac{t^\alpha}{\Gamma(1+\alpha)}.
\]
Write $D$ for the space of right-continuous
paths with left limits with the Skorokhod $J_1$ topology, and $M_p$ for
locally finite point measures on $[0,T]\times\spc$ with the vague topology.
Constants in estimates may depend on the fixed model parameters and $T$.

\section{Main Results}\label{sec:main}

We construct the order flow model using a marked INAR process. 
Then, we identify its joint limit as the rough Hawkes--Heston model. 
Section~\ref{sec:option:prices} then applies the log-price convergence to
European options and contracts monitored over the life of the option.

\subsection{Marked INAR model}\label{sec:model}

Fix $1/2<\alpha<1$, $v_0,\eta\ge0$, $b<0$, $c,\xi>0$ and
$\Lambda\ge0$. Let $\nu$ be a finite positive measure on $\spc$ with
\[
 m_j:=\int_{\spc}z^j\nu(\dd z)<\infty,\qquad j=0,1,2.
\]
Here $c$ is the variance of the continuous variance driver per unit activity.
Its diffusion coefficient is $\sqrt c$. Specify the lag law by its tail
\begin{equation}\label{eq:tail}
\begin{aligned}
 s_0&:=1, & s_n&:=\frac{n^{-\alpha}}{\Gamma(1-\alpha)},\quad n\ge1,\\
 \phi_0&:=0, & \phi_k&:=s_{k-1}-s_k,\quad k\ge1.
\end{aligned}
\end{equation}
This is the lattice law of \citet[Definition~2.1]{wcz}. Choose a sign asymmetry
$\beta>1$ and a relative up-tick $0<q\le1$. Define
\begin{equation}\label{eq:weights}
\begin{gathered}
 \omega_+:=\frac{2}{1+q},\qquad \omega_-:=\frac{2q}{1+q},\qquad
 (\gamma_+,\gamma_-):=(1,\beta),\qquad (p_+,p_-):=(q,-1),\\
 D_0:=\sum_{i\in\sig}\omega_i\gamma_i,\qquad
 H:=\sum_{i\in\sig}\omega_i p_i^2,\\
 \ell_0:=\frac{\sum_{i\in\sig}\omega_i\gamma_i^2}{cD_0^2},\qquad
 \ell:=\ell_0h^{1-2\alpha},\qquad r_i:=\omega_i\ell,
\end{gathered}
\end{equation}
and let
\begin{equation}\label{eq:loads}
\begin{gathered}
 y_i:=p_i\sqrt{\frac{h}{\ell H}},\quad
 g_i:=\frac{h^{1-\alpha}\gamma_i}{\ell D_0},\quad
 d:=\xi m_1-b,\quad a:=(1+d h^\alpha)^{-1}.
\end{gathered}
\end{equation}
Let $\F_n$ contain all outcomes through bin $n-1$. The activity $V_n$ in
bin $n$ is $\F_n$-measurable. Conditional on $\F_n$, the two regular
counts 
\[
N_{n,i}\sim{\rm Pois}(r_iV_n),
\]
and the Poisson random measure
$\Pi_n$ with mean $hV_n\nu(\dd z)$ are mutually independent. Let
\[
 Z_n:=\int_{\spc}z\Pi_n(\dd z).
\]
The
predictable recursion, with empty history and $n=0,\ldots,N-1$, is
\begin{equation}\label{eq:model}
\begin{aligned}
 H_n&:=a\left(v_0s_n+\eta h^\alpha\right),\\
 Y_n&:=a h^{\alpha-1}\left(\sum_{i\in\sig}g_iN_{n,i}+\xi Z_n\right),\\
 V_n&:=H_n+\sum_{j=0}^{n-1}\phi_{n-j}Y_j.
\end{aligned}
\end{equation}
In particular $V_0=H_0$ on the grid and $V_0\to v_0$. Use the same aggregate
mark in the log-price increment, with $X^h_0=0$, $S^h=S_0\exp X^h$,
\begin{equation}\label{eq:price}
\begin{aligned}
 D_h&=\sum_{i\in\sig}r_i\left(\e^{y_i}-1\right)
       +h\int_{\spc}\left(\e^{-\Lambda z}-1\right)\nu(\dd z),\\
 \Delta X_n&=\sum_{i\in\sig} y_iN_{n,i}-\Lambda Z_n-D_hV_n,\\
 X^h_t&=\sum_{\substack{0\le n<N\\t_{n+1}\le t}}\Delta X_n.
\end{aligned}
\end{equation}
The price changes at the right endpoint of the bin, and its mark first changes
activity in the next bin. Multiple marks within one bin are all retained.
The regular price innovation has conditional variance $hV_n$. Including
the conditionally independent mark sum gives
$\operatorname{Var}(\Delta X_n\mid\F_n)=hV_n(1+\Lambda^2m_2)$.

Let $E:=\sig\cup\spc$ be the event type space, where $+$ and $-$ label the
two regular event types. Extend $\nu$ by zero on $\sig$. Define the mean
arrival measure per unit activity and the excitation weight of a parent of
type $e\in E$ by
\begin{equation}\label{eq:offspring-weights}
\begin{gathered}
 \mu_h:=r_+\delta_++r_-\delta_-+h\nu,\\
 w_h(i):=a h^{\alpha-1}g_i\quad(i\in\sig),\qquad
 w_h(z):=a h^{\alpha-1}\xi z\quad(z\in\spc).
\end{gathered}
\end{equation}
Write $\widehat N_{n,i}:=N_{n,i}-r_iV_n$ for the centered regular counts.
The next proposition computes the mean offspring operator from
\eqref{eq:offspring-weights} and the covariance of the regular innovations
from the loadings in \eqref{eq:loads}.

\begin{proposition}\label{prop:model}
The activity recursion \eqref{eq:model} is nonnegative. The mean offspring
operator, summed over all lags, is
\begin{equation}\label{eq:mean-offspring-operator}
 (\mathcal M_h f)(e)=w_h(e)\int_E f(e')\mu_h(\dd e'),
 \qquad f\in L^2(E,\mu_h),\quad \mu_h\text{-a.e. }e\in E.
\end{equation}
Its spectral radius is $a(1+\xi m_1h^\alpha)<1$. The discounted stock
$S^h$ is a martingale. The two regular innovations have conditional
covariance matrix $hV_n Q$, i.e. 
\begin{equation}\label{eq:cov}
 \operatorname{Cov}\!\left(
     \sum_{i\in\sig}(g_i,y_i)^\top\widehat N_{n,i}\,\middle|\,\F_n\right)
   =V_n\sum_{i\in\sig}r_i(g_i,y_i)^\top(g_i,y_i)=hV_nQ,
\end{equation}
where $hV_n$ is the bin's increment of
activity time and
\begin{equation}
 Q=\begin{pmatrix}c&\rho\sqrt c\\ \rho\sqrt c&1\end{pmatrix},\qquad
 \rho=-\frac{(\beta-1)\sqrt q}{\sqrt{(1+q)(1+\beta^2q)}}.
\end{equation}
Every $\rho\in(-1,0)$ is attainable. For $m_0>0$, as $h\downarrow0$ the
marked share of the arrival rate per unit activity and the offspring
weight of a fixed mark $z>0$ relative to a regular parent of type $i\in\sig$
satisfy
\begin{equation}\label{eq:rare-mark-scaling}
\begin{aligned}
 \frac{\mu_h(\spc)}{\mu_h(E)}
 &=\frac{hm_0}{2\ell+hm_0}
   =\frac{m_0h^{2\alpha}}{2\ell_0+m_0h^{2\alpha}}
   \sim \frac{m_0}{2\ell_0}h^{2\alpha},\\
 \frac{w_h(z)}{w_h(i)}
 &=\frac{\xi z}{g_i}
   =\frac{\xi z\ell_0D_0}{\gamma_i}h^{-\alpha}.
\end{aligned}
\end{equation}
\end{proposition}

\begin{proof}
We provide the proof in Appendix~\ref{app:model-proof}.
\end{proof}

We use regular martingales $B^h,W^h$,
the cumulative number and mass of marked events $C^h,J^h$, and a predictable
activity clock $A^h$. They all start at zero and are constant between bin endpoints,
with increments
\begin{equation}\label{eq:drivers}
\begin{aligned}
 \left(\Delta B_n,\Delta W_n\right)&=\sum_{i\in\sig}(g_i,y_i)\widehat N_{n,i},\\
 \left(\Delta C_n,\Delta J_n\right)&=\left(\Pi_n(\spc),Z_n\right),\\
 \Delta A_n&=hV_n,\\
 \Delta M_n&=\Delta B_n+\xi\left(Z_n-m_1 hV_n\right).
\end{aligned}
\end{equation}
For the point measure with marks retained let
\[
 \J^h(\dd t,\dd z):=\sum_{n=0}^{N-1}\delta_{t_{n+1}}(\dd t)\Pi_n(\dd z).
\]
Let $V^h(t):=V_{\lfloor t/h\rfloor}$ on $[0,T)$ and $V^h(T):=V_{N-1}$.
The step function defines the same element of $L^2(0,T)$ for every choice
of $V^h(T)$, since $\{T\}$ has Lebesgue measure zero.

To express the grid recursion in the form of the limiting variance equation, let
\[
 U_0:=1,\qquad U_n:=\sum_{j=1}^n\phi_j U_{n-j},\qquad
 u_n:=U_n-\ind_{\{n=0\}}.
\]
For $0\le t<T$, the deterministic part of the grid variance and its limit are
\[
 G_h(t)=a\left(v_0+\eta h^\alpha\sum_{j=0}^{\lfloor t/h\rfloor}U_j\right),
 \qquad G_0(t)=v_0+\eta\G(t).
\]
The renewal sequence $U$ and the compensated increments in \eqref{eq:drivers} give, for $0\le n<N$,
\begin{equation}\label{eq:mild}
 V_n=G_h(t_n)+a\sum_{j=0}^{n-1}h^{\alpha-1}u_{n-j}
                         \left(b h V_j+\Delta M_j\right).
\end{equation}
Lemma~\ref{lem:renew} in Appendix~\ref{app:identification} derives \eqref{eq:mild} from \eqref{eq:model}.

\subsection{The joint rough Hawkes--Heston limit}
Let $V$ be a random element of $L^2(0,T)$ that is nonnegative almost
everywhere, with continuous adapted integral
\[
 A_t:=\int_0^t V_s\dd s.
\]
In stochastic integrals and compensators we use the predictable version of $V$
given by the finite limit of $n(A_t-A_{(t-1/n)^+})$ where it exists, and
zero elsewhere. The Lebesgue differentiation theorem identifies this version
with $V$ almost everywhere. Let $B,W$ be continuous
local martingales starting at zero, with bracket matrix $QA$ in the order $B,W$.
Let $\J$ be a point measure with no mark at time zero, simple in time, with compensator
$V_s\dd s\,\nu(\dd z)$ in the same filtration. Let
\begin{gather*}
 J_t:=\int_{(0,t]\times\spc}z\J(\dd s,\dd z),\qquad
 C_t:=\J((0,t]\times\spc),\\
 M_t:=B_t+\xi\left(J_t-m_1A_t\right),\qquad
 \kappa:=\frac12+\int_{\spc}\left(\e^{-\Lambda z}-1\right)\nu(\dd z).
\end{gather*}
The target system is
\begin{align}
 V(t)&=G_0(t)+\int_0^t\K(t-s)bV_s\dd s+
                         \int_{[0,t)}\K(t-s)\dd M_s, \label{eq:target}\\
 X_t&=W_t-\Lambda J_t-\kappa A_t,\qquad S_t=S_0\e^{X_t}. \label{eq:tprice}
\end{align}
The variance equation \eqref{eq:target} is an equality in $L^2(0,T)$.
Its convolution integrals are defined for Lebesgue-a.e.\ $t\in(0,T)$.
In the Brownian representation the continuous coefficients are
$\sqrt{cV_t}$ and $\sqrt{V_t}$ with correlation $\rho$.
Indeed, normalize the martingales
\[
 \frac{B}{\sqrt c},\qquad
 \frac{W-\rho B/\sqrt c}{\sqrt{1-\rho^2}}
\]
by integrating $V^{-1/2}$
on $\{V>0\}$. On $\{V=0\}$ fill in the two missing brackets with an
independent two-dimensional Brownian motion on an extension. The resulting
continuous martingales have bracket $tI_2$, so L\'evy's characterization
gives the two Brownian drivers. It\^o's formula in \eqref{eq:tprice} gives
\[
 \frac{\dd S_t}{S_{t-}}=\dd W_t+
   \int_{\spc}\left(\e^{-\Lambda z}-1\right)
       \left(\J(\dd t,\dd z)-V_t\nu(\dd z)\dd t\right),
\]
since the continuous quadratic variation contributes $\frac12\dd A_t$
and a raw mark multiplies the stock by $\e^{-\Lambda z}$.
Under the change of mark $y=\xi z$, the mark measure in the compensator becomes
the pushforward of $\nu$ by $z\mapsto\xi z$. In the notation of \citet{bps},
the variance driver jumps by $y$ and the log price by $-(\Lambda/\xi)y$.

Following \citet[Proposition~2.3]{wcz}, define the normalized counts and
centered martingales for the two regular event types $i\in\sig$ by
\begin{equation}\label{eq:regular-scaled}
\begin{aligned}
 L^{h,i}_t&:=\frac h{r_i}\sum_{\substack{0\le n<N\\t_{n+1}\le t}}N_{n,i},\\
 Z^{h,i}_t&:=\sqrt{\frac h{r_i}}\sum_{\substack{0\le n<N\\t_{n+1}\le t}}\widehat N_{n,i},
 \qquad {\bf Z}^h:=\left(Z^{h,+},Z^{h,-}\right)^\top.
\end{aligned}
\end{equation}
Equations~\eqref{eq:drivers} and \eqref{eq:regular-scaled} give
$(B^h,W^h)^\top=\mathsf P{\bf Z}^h$, where
\begin{equation}\label{eq:regular-map}
 \mathsf P_{\cdot i}:=\sqrt{r_i/h}\,(g_i,y_i)^\top
 =\left(\frac{\gamma_i\sqrt{\omega_i}}{D_0\sqrt{\ell_0}},
         \frac{p_i\sqrt{\omega_i}}{\sqrt H}\right)^\top,\qquad i\in\sig.
\end{equation}
This matrix is independent of $h$, and $\mathsf P\mathsf P^\top=Q$ by \eqref{eq:cov}.
In particular $\mathsf P$ is invertible since $|\rho|<1$.

The following theorem identifies the joint limit of the grid variance,
regular martingales, log price and marked point measure as the system
\eqref{eq:target}--\eqref{eq:tprice}.
It also shows that both normalized regular counts converge to the same
activity clock, while their centered fluctuations retain the covariance
structure in \eqref{eq:regular-map}.
\begin{theorem}[Joint scaling limit]\label{thm:limit}
For the parameters of \eqref{eq:tail}--\eqref{eq:loads},
\[
 \left(V^h,B^h,W^h,X^h,\J^h\right)
 \ \Longrightarrow\ (V,B,W,X,\J)
\]
in the product space
\[
 L^2(0,T)\times D([0,T],\R^3)\times M_p.
\]
The limiting joint law solving \eqref{eq:target}--\eqref{eq:tprice} is unique.
The clock $A^h$ converges jointly to $A$ in $D$ with continuous limit.
The two counts in \eqref{eq:regular-scaled} both converge jointly to $A$.
The regular martingale vector converges jointly to
${\bf Z}=\mathsf P^{-1}(B,W)^\top$, with $\mathsf P$ as in
\eqref{eq:regular-map} and bracket $I_2A$ in the joint filtration.
\end{theorem}
\begin{proof}
We provide the proof in Appendix~\ref{app:joint-limit-proof}.
\end{proof}

%


\section{Option prices}\label{sec:option:prices}

\subsection{European option prices}\label{sec:European:prices}

The terminal-value map is continuous for the $J_1$ topology, so
Theorem~\ref{thm:limit} gives convergence of European put option prices.
The martingale property of the discounted stock in both models then gives
convergence of European call option prices through put--call parity.

\begin{corollary}[European option prices]\label{cor:european-prices}
Under the assumptions of Theorem~\ref{thm:limit}, for every strike $K>0$ and maturity $T>0$,
\begin{equation}\label{eq:european-price-convergence}
\begin{aligned}
 \lim_{h\downarrow0}\E\left[\left(K-S^h_T\right)^+\right]
   &=\E\left[\left(K-S_T\right)^+\right],\\
 \lim_{h\downarrow0}\E\left[\left(S^h_T-K\right)^+\right]
   &=\E\left[\left(S_T-K\right)^+\right].
\end{aligned}
\end{equation}
\end{corollary}

\begin{proof}
We provide the proof in Appendix~\ref{app:european-prices}.
The argument also verifies the identity
$\E[S_t]=S_0$ used below for arithmetic Asian call options.
\end{proof}

\subsection{Path-dependent option prices}\label{sec:path-prices}

Let $S_0>0$ and let $K>0$ be the strike in discounted stock units.
For a positive c\`adl\`ag price path $s$ on $[0,T]$, define its arithmetic
average and running minimum by
\[
 \overline s_T:=\frac1T\int_0^T s(t)\dd t,\qquad
 \underline s_T:=\inf_{0\le t\le T}s(t).
\]
With the price convention in \eqref{eq:price}, the corresponding quantities for
the marked INAR model are exactly
\begin{equation}\label{eq:grid-path-payoffs}
 \overline S^h_T:=\frac1N\sum_{n=0}^{N-1}S^h_{t_n},\qquad
 \underline S^h_T:=\min_{0\le n\le N} S^h_{t_n}.
\end{equation}
The price at $T$ enters the minimum. It receives zero weight in the time
average because each return is recorded at the right endpoint of its bin.
We monitor a lower barrier at every grid date, so that the monitoring mesh
also tends to zero in the continuous-time limit.

The payoff of a down-and-out option can be discontinuous when the minimum equals the barrier.
To rule out an atom below the initial price, we condition on the variance,
its continuous driver and the marks, then use the remaining Gaussian component
of the log price.

\begin{lemma}\label{lem:barrier-atom}
Under the assumptions of Theorem~\ref{thm:limit}, for every $0<L<S_0$,
\[
 \Prob(\underline S_T=L)=0 .
\]
\end{lemma}
\begin{proof}
We provide the proof in Appendix~\ref{app:barrier-atom-proof}.
\end{proof}

Next, we study the convergence of path-dependent option prices under marked INAR model
to that under rough Hawkes--Heston model. 
The Asian option payoffs depend on the time average. The down-and-out put has zero
rebate and is knocked out on touching the barrier.
Their payoff functions are summarized in Table~\ref{tab:payoff}.

\begin{table}
\begin{center}
\begin{tabular}{ll}
\toprule
Contract & $F(s)$\\
\midrule
Arithmetic Asian put option & $(K-\overline s_T)^+$\\
Arithmetic Asian call option  & $(\overline s_T-K)^+$\\
Down-and-out put option & $(K-s(T))^+\ind_{\{\underline s_T>L\}}$\\
\bottomrule
\end{tabular}
\end{center}
\caption{Payoff functions for path-dependent options.}
\label{tab:payoff}
\end{table}

The next corollary combines Theorem~\ref{thm:limit} and
Lemma~\ref{lem:barrier-atom} to justify convergence of the grid prices
for all three payoffs in Table~\ref{tab:payoff}.

\begin{corollary}[Path-dependent option prices]\label{cor:path-prices}
Under the assumptions of Theorem~\ref{thm:limit}, fix the strike $K>0$ and the lower barrier $0<L<S_0$.
For each payoff $F$ in Table~\ref{tab:payoff},
\begin{equation}\label{eq:path-price-convergence}
 \lim_{h\downarrow0}\E\left[F\left(S^h\right)\right]=\E[F(S)].
\end{equation}
\end{corollary}
\begin{proof}
We provide the proof in Appendix~\ref{app:path-prices-proof}.
\end{proof}

\section{Numerical experiments}\label{sec:num}

Both experiments have maturity $T=1$, unit forward $F=S_0=1$, and
INAR grids with $N=128,512,2{,}048$ bins.
The macroscopic parameters other than roughness follow \citet[Table~1]{bps},
with unit mark loading in the variance driver and exponential marks,
$b=-1.812$, $c=0.115$, $\rho=-0.731$, $\Lambda=0.276$, $v_0=0.0079$, $\eta=0.049$,
$\xi=1$, and $\nu(\dd z)=\e^{-z}\dd z.$
For each $\alpha$, we set $\beta=8$ and solve
\eqref{eq:cov} on $0<q<1/8$, giving $q\simeq0.041605$.

\subsection{One-year European option implied volatility smiles}\label{subsec:smiles}

\begin{figure}[!tbp]
\centering
\includegraphics[width=\linewidth]{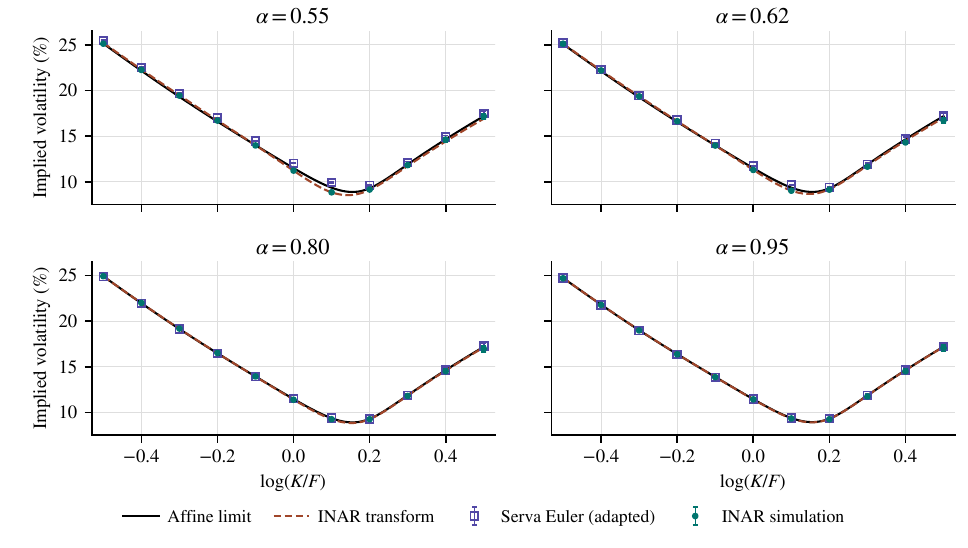}
\caption{One-year smiles for the exponential mark law. The transform curves use
101 log-forward strikes. The marked INAR simulation and the adapted Serva Euler
scheme show Monte Carlo estimates and pointwise 95\% intervals at eleven
strikes. Both simulations and the INAR transform use 512 time steps.}\label{fig:smiles}
\end{figure}

Figure~\ref{fig:smiles} compares one-year European option smiles from the marked
INAR model and an adaptation of the multifactor Euler scheme of \citet{serva}
with the rough Hawkes--Heston model reference.
Following \citet[Figure~1]{wcz}, we vary $\alpha$ over $0.55,0.62,0.80,0.95$.
At each value, the discrete
variance equation uses $G_h$ from \eqref{eq:mild}, which converges
uniformly to the initial curve $G_0(t)=v_0+\eta\G(t)$.

For the finite-grid curve let $h:=1/N$, $w:=1/2+\ii u$, and
$\Phi_h(w):=\E[\exp(wX^h_1)]$. Conditioning backward in
\eqref{eq:model}--\eqref{eq:price} gives the recursion, for $0\le n<N$,
\begin{align}
 x_n&=a h^{\alpha-1}\sum_{j=0}^{n-1}\phi_{n-j}U^\star_j,\nonumber\\
 U^\star_n&=\sum_{i\in\sig}r_i\left(\e^{w y_i+x_ng_i}-1\right)
       +h\left(\frac{1}{1+\Lambda w-\xi x_n}-1\right)-wD_h,\nonumber\\
 \Phi_h(w)&=\exp\!\left(\sum_{n=0}^{N-1}H_{N-1-n}U^\star_n\right).
 \label{eq:grid-transform}
\end{align}
Indeed, when integrating out bin $m=N-1-n$, the later bins contribute
the coefficient $s_n^\star=\sum_{j=0}^{n-1}\phi_{n-j}U_j^\star$ of $Y_m$.
With $x_n=a h^{\alpha-1}s_n^\star$, the conditional Poisson formula gives
\[
 \E\!\left[\exp\left(w\Delta X_m+s_n^\star Y_m\right)\mid\F_m\right]=\exp\left(V_m U_n^\star\right).
\]
Substitution of $V_m$ from \eqref{eq:model} retains the constant
$H_mU_n^\star$ and adds $\phi_{m-l}U_n^\star$ to the coefficient
of each earlier $Y_l$, $0\le l<m$. Backward induction gives the last
line of \eqref{eq:grid-transform}.

The mark integral in each conditioning step is finite on this contour.
Indeed $x_0=0$. If $\mathrm{Re}\,x_n\le0$, then
$\mathrm{Re}(1+\Lambda w-\xi x_n)\ge1$.
Since $\mathrm{Re}\,w=1/2$ and $\e^{v/2}-1\le(\e^v-1)/2$ for real $v$,
\eqref{eq:price} gives
\[
 \mathrm{Re}\,U_n^\star\le
 \sum_{i\in\sig}r_i\left(\e^{y_i/2}-1\right)
 +h\int_{\spc}\left(\e^{-\Lambda z/2}-1\right)\nu(\dd z)-D_h/2\le0 .
\]
The nonnegative lag weights now give $\mathrm{Re}\,x_{n+1}\le0$ by induction.
For the continuous curve we evaluate $\Phi(w)=\E[\exp(wX_1)]$
using the Riccati--Volterra transform of \citet[Theorems~3.1 and~3.3]{bps}.
We approximate the fractional kernel by a sum of 129 exponentials, including
one constant term, and use linear product integration on $1{,}024$ time
intervals. For $\Psi=\Phi_h$ or $\Phi$ and $k=\log(K/F)\in\R$,
the normalized call option price is given by the inversion formula of
\citet[Proposition~3.5]{bps},
\[
 \frac{C(1,K)}{S_0}=1-\frac{\e^{k/2}}{\pi}
 \int_0^\infty
 \frac{\mathrm{Re}\!\left[\e^{-\ii u k}\Psi(1/2+\ii u)\right]}
      {u^2+1/4}\dd u .
\]
To see that the formula applies at every strike in both models, write $x$
for the terminal log return. Then
$\min(\e^x,\e^k)=\e^{(x+k)/2}\e^{-|x-k|/2}$.
Fourier inversion of $\e^{-|y|/2}$ gives
$\pi^{-1}\int_0^\infty\cos(uy)(u^2+1/4)^{-1}\dd u=\e^{-|y|/2}$.
Since $\E[\e^x]=1$ and $\E[\e^{x/2}]\le1$, Fubini's theorem applies
when taking expectations in $(\e^x-\e^k)^+=\e^x-\min(\e^x,\e^k)$.
We truncate at $u=480$ and apply Gauss--Legendre quadrature on intervals
of width four, with 32 nodes on the first interval and 16 on each subsequent
interval.

Figure~\ref{fig:smiles} plots the
$512$-bin and continuous transform curves at 101 equally spaced log-forward
strikes on $[-0.5,0.5]$. To check the $512$-bin curve, we simulate
$2^{20}=1{,}048{,}576$ independent paths on this grid for each $\alpha$.
The same paths price all eleven strikes
$k=-0.5,-0.4,\ldots,0.5$ in each panel. We use put options for $k<0$ and call options
otherwise. To check the second moment on this grid, let $R_h(x)$ denote the
right-hand side defining $U_n^\star$ in \eqref{eq:grid-transform} at $w=2$
and $x_n=x$. It is increasing for $x<1+2\Lambda$, since $\xi=1$.
For each of the four parameter sets, substitution of \eqref{eq:loads} gives
$R_h(1)<h^{1-\alpha}/a$. Starting at $x_0=0$, the nonnegative lag
weights with sum at most one therefore give $x_n\le1$ at every step.
The mark denominator is at least $2\Lambda>0$, so $\Phi_h(2)<\infty$
and the payoffs have finite variance. We invert the Black price at the sample mean and at
the endpoints of its normal interval, $1.96$ estimated standard errors on
either side, to obtain a pointwise asymptotic 95\% interval for implied volatility.
The $512$-bin transform value falls within the interval at 42 of the
44 strike--parameter pairs. The two exceptions occur at $\alpha=0.62$
and $k\in\{0.4,0.5\}$.

The multifactor Euler comparison uses the same parameters, $512$ steps and
$2^{20}$ independent paths for each value of $\alpha$.
We use $64$ exponential factors from the geometric quadrature of
\citet[Section~2.1]{serva} and add a constant factor for the kernel mass
below the first quadrature interval. On each time step we freeze the
continuous contribution to the jump intensity and use the left-endpoint
intensity as the initial thinning bound. The jump factors decay between
proposals and are updated after each accepted mark. We use the variance
floor $10^{-5}$ of \citet[Algorithm~A.1]{serva}.
Let $V_n^{\mathrm E}$ denote the floored left-endpoint variance,
$Z_n^{\mathrm E}$ the total size of the marks accepted during the step and
$(G_n^V,G_n^S)$ a standard normal pair with correlation $\rho$.
The Euler updates use the diffusion coefficients and jump compensation
of the continuous-time equations. With $m_j$ and $x_j$ the weight and
decay rate of factor $j$, they are
\[
\begin{aligned}
 C_{j,n+1}
 &=\e^{-x_j h}C_{j,n}
   +m_j\left((b-1)V_n^{\mathrm E}h+
             \sqrt{cV_n^{\mathrm E}h}\,G_n^V\right),\\
 \Delta X_n^{\mathrm E}
 &=\left(-\frac12+\frac{\Lambda}{1+\Lambda}\right)V_n^{\mathrm E}h
   +\sqrt{V_n^{\mathrm E}h}\,G_n^S-\Lambda Z_n^{\mathrm E}.
\end{aligned}
\]
Here $C_{j,n}$ is the continuous contribution to factor $j$. The same
marks enter the jump factors and the log price. We invert the Monte Carlo
prices and their pointwise intervals at the same eleven strikes as for the
INAR simulation. The maximum absolute differences between the Euler estimates
and the affine reference are $0.564$, $0.266$, $0.076$ and $0.039$ volatility
percentage points in increasing order of $\alpha$. The variance floor binds at
$44.82\%$, $34.82\%$, $10.27\%$ and $1.57\%$ of grid updates, respectively.

Table~\ref{tab:grid} compares the lattice transform with the continuous
reference at the same eleven strikes. For each $\alpha$, the maximum
absolute implied-volatility difference decreases with each grid refinement.
On all three grids the maximum occurs at $k=0.1$ for all four values of
$\alpha$.

\begin{table}[!tbp]
\centering
\caption{Maximum absolute implied-volatility difference between the lattice
transform and the continuous affine reference over the eleven one-year strikes,
in volatility percentage points.}\label{tab:grid}
\begin{tabular}{lrrr}
\toprule
$\alpha$&128 steps&512 steps&$2{,}048$ steps\\
\midrule
0.55&1.119&0.510&0.234\\
0.62&0.790&0.326&0.137\\
0.80&0.322&0.106&0.035\\
0.95&0.153&0.041&0.011\\
\bottomrule
\end{tabular}
\end{table}

\FloatBarrier
\subsection{Arithmetic Asian and down-and-out put options}\label{subsec:path-numerics}

We price an arithmetic Asian put option and a down-and-out put option with strike $K=1$
at $\alpha=0.62$ and $0.80$. The down-and-out barrier is $L=0.85$.
At $K=S_0$, Asian option put--call parity gives identical Asian call and put option prices, so we report the put.

\paragraph{Simulation by block convolution.}
The regular counts and the common marks enter the same scalar excitation $Y_j$
in \eqref{eq:model}. Thus the history needed before drawing bin $n$ is the
one-sided convolution $\sum_{j=0}^{n-1}\phi_{n-j}Y_j$. We adapt the
divide-and-conquer FFT algorithm of \citet[Section~4.1, Algorithm~1]{wcz}
to this marked excitation. Consider a completed block of $d$ bins starting at
$m$, followed by a block of at most $d$ bins. Its contribution to the activity
in bin $m+d+k<N$ of the second block is
\begin{equation}\label{eq:fft-block-history}
 \sum_{i=0}^{d-1}\phi_{d+k-i}Y_{m+i}
 =\left[\operatorname{IFFT}_{2d}\left(
       \operatorname{FFT}_{2d}(y)\odot\operatorname{FFT}_{2d}(p)
   \right)\right]_{d+k},\qquad 0\le k<d ,
\end{equation}
where $y=(Y_m,\ldots,Y_{m+d-1},0,\ldots,0)$ and
$p=(\phi_0,\ldots,\phi_{2d-1})$ have length $2d$, with $\phi_0=0$.
Kernel entries beyond the simulation horizon may be set to zero, and
$\odot$ denotes componentwise multiplication.
Since $1\le d+k-i\le 2d-1$, the retained entries give the linear
convolution in \eqref{eq:fft-block-history}.

We precompute the kernel transform for each block length. Within blocks of
$64$ bins or fewer, we sum the lags directly and draw the counts in time
order. For the exponential mark law, a bin with $j\ge1$ marks has total
mark size $Z_n$ distributed as Gamma with shape $j$ and unit scale.
The same $Z_n$ enters $Y_n$ and the return in \eqref{eq:price}.
Every pair of bins in distinct small blocks contributes once, when its earlier
bin lies in the left half and its later bin in the right half of their smallest
common block. At each block length $2d$ there are at most
$\lceil N/(2d)\rceil$ block updates per path, each costing $\mathcal O(d\log d)$.
Summing over the dyadic lengths gives $\mathcal O(N\log^2 N)$ operations per path
to compute the excitation history. Along each path we update the arithmetic
average and running minimum in \eqref{eq:grid-path-payoffs}. We estimate the
Asian and down-and-out put option prices by their respective sample means.

\paragraph{Grid refinement.}
Table~\ref{tab:path-refinement} compares $N=128,512,2{,}048$ bins over one
year. A separate $N=16{,}384$ estimate provides a finite-grid reference.
Each combination of $N$ and $\alpha$ uses $200{,}000$ independent
paths, with independent random streams across combinations. Prices are
reported per $100$ units of spot. For each price, parentheses give the
sample standard deviation of the payoff divided by $\sqrt{200{,}000}$.
For a difference between independent grid estimates, the standard error is
the square root of the sum of their squared standard errors.
For each of these three grids and each $\alpha$, timing starts before kernel setup
and ends after both payoffs have been averaged over the same $200{,}000$
simulated paths. We report the median elapsed time of three runs with
identical seeds after compilation, using ten worker threads on an Apple M4 Pro.

\begin{table}[!htbp]
 \centering
 \caption{Path-dependent put option prices per $100$ units of spot.
 The final column gives the $2{,}048$-bin price minus the reference, in
 basis points of spot. Parentheses give Monte Carlo standard errors in the
 units of the corresponding column. Each runtime covers all $200{,}000$ paths and both
 put option prices for one $(\alpha,N)$ pair.}
 \label{tab:path-refinement}
 \small
 \setlength{\tabcolsep}{4pt}
 \begin{tabular}{@{}clrrrrr@{}}
 \toprule
 & & \multicolumn{3}{c}{Simulation grid} & Reference & Difference\\
 \cmidrule(lr){3-5}
 $\alpha$ & Quantity & $128$ & $512$ & $2{,}048$ & $16{,}384$ & (bp)\\
 \midrule
0.62 & Arithmetic Asian put option & $2.360\ (0.011)$ & $2.411\ (0.011)$ & $2.442\ (0.011)$ & $2.446\ (0.011)$ & $-0.4\ (1.6)$\\
 & Down-and-out put option & $0.969\ (0.005)$ & $0.950\ (0.005)$ & $0.938\ (0.005)$ & $0.914\ (0.005)$ & $+2.4\ (0.7)$\\
 & Total runtime (s) & 0.50 & 1.98 & 9.76 & --- & ---\\
\addlinespace
0.80 & Arithmetic Asian put option & $2.383\ (0.011)$ & $2.447\ (0.011)$ & $2.443\ (0.011)$ & $2.431\ (0.011)$ & $+1.2\ (1.6)$\\
 & Down-and-out put option & $1.056\ (0.006)$ & $1.032\ (0.006)$ & $0.984\ (0.005)$ & $0.968\ (0.005)$ & $+1.6\ (0.8)$\\
 & Total runtime (s) & 0.53 & 2.06 & 10.52 & --- & ---\\
 \bottomrule
 \end{tabular}
\end{table}

Moving from $512$ to $2{,}048$ bins changes the four estimated prices by
$0.4$--$4.8$ basis points of spot in absolute value. Relative to the
$16{,}384$-bin reference, the $2{,}048$-bin Asian put option estimate lies
$0.4$ basis points below at $\alpha=0.62$ and $1.2$ basis points above
at $\alpha=0.80$. Both differences are smaller than one Monte Carlo standard
error. The down-and-out estimates exceed their respective references by
$2.4$ and $1.6$ basis points of spot, about $3.3$ and $2.2$ standard
errors, respectively. For the barrier
option contract, refinement changes both the marked INAR price dynamics and the set of
dates on which a knockout can occur. The fixed-strike lookback put shows
larger differences in these experiments, with $2{,}048$-bin estimates
$29.3$ and $17.7$ basis points of spot below the respective references.

\section{Conclusion}\label{sec:end}

We propose the marked INAR model as an implementable discrete-time Monte Carlo algorithm for
the rough Hawkes--Heston model. We provide
rigorous convergence results of the discrete-time marked INAR model
to the continuous-time rough Hawkes--Heston model, which provides
a solid theoretical foundation for the proposed algorithm.

We implement the algorithm with a
divide-and-conquer FFT scheme that computes the excitation history in
$\mathcal O(N\log^2 N)$ operations per path on an $N$-bin grid.
As the grid is refined, the log price, spot variance and common marked point
measure converge jointly. With positive price-jump loading, the same marked
order lowers the stock price and excites future variance. In the European
option experiment, the one-year implied volatility smile on the INAR grid
moved closer to the continuous-model smile at each refinement for all four
roughness levels tested.

The convergence of price paths also justifies using the marked INAR model for
contracts that might depend on the stock's history, including
the path-dependent options such as arithmetic Asian options and barrier options. 
As the monitoring grid is
refined, arithmetic Asian call and put options and down-and-out put options converge
in price to their continuously monitored counterparts. The barrier option result
holds for every positive barrier below the initial stock price. The Monte Carlo experiments
are presented to demonstrate the efficiency of our numerical scheme.

\appendix


\section{Technical Proofs}\label{app:proofs}

\subsection{Proof of Proposition~\ref{prop:model}}\label{app:model-proof}

\begin{proof}
In \eqref{eq:tail}, $s_1<1$, $s_n\downarrow0$, and hence
\[
 \phi_k\ge0,\qquad \sum_{k=1}^\infty\phi_k=1.
\]
By \eqref{eq:weights}--\eqref{eq:model},
\[
 \sum_{i\in\sig}r_i g_i=h^{1-\alpha},\qquad
 \E[Y_n\mid\F_n]=a\left(1+\xi m_1 h^\alpha\right)V_n.
\]
In the thinning construction, a parent of type $e$ generates an
independent Poisson offspring measure at lag $k\ge1$ with mean
$\phi_k w_h(e)\mu_h$, using \eqref{eq:offspring-weights}.
Independent Poisson immigration in bin $n$ has mean measure
$H_n\mu_h$. Superposition gives the conditional mean measure
$V_n\mu_h$ of \eqref{eq:model}. For a test function
$f\in L^2(E,\mu_h)$, the expected sum of $f$ over the offspring
of a type-$e$ parent is
\[
 \sum_{k=1}^\infty \phi_k w_h(e)\int_E f(e')\mu_h(\dd e')
 =w_h(e)\int_E f(e')\mu_h(\dd e'),
\]
which proves \eqref{eq:mean-offspring-operator}.
Since $\mu_h(E)<\infty$ and $w_h\in L^2(E,\mu_h)$ by $m_2<\infty$,
the operator is bounded and has rank one. Its unique nonzero eigenvalue is
\[
 \int_E w_h(e)\mu_h(\dd e)
 =a h^{\alpha-1}\left(\sum_{i\in\sig}r_i g_i+\xi h m_1\right)
 =a(1+\xi m_1h^\alpha).
\]
By the definition of $a$ in \eqref{eq:loads}, its gap from one is
\[
 1-a(1+\xi m_1h^\alpha)=-ab h^\alpha>0.
\]
Induction in \eqref{eq:model} gives nonnegative, finite bin intensities.
The conditional Poisson exponential formula applied to \eqref{eq:price} gives
\[
 \E[\exp(\Delta X_n)\mid\F_n]=\exp\{V_n(D_h-D_h)\}=1.
\]
By conditional independence and the Poisson variance formula,
\[
 \operatorname{Cov}\!\left(\widehat N_{n,i},\widehat N_{n,j}\mid\F_n\right)
 =\ind_{\{i=j\}}r_iV_n,\qquad i,j\in\sig .
\]
Thus, the covariance in \eqref{eq:cov} is $V_n$ times the sum of the two
weighted outer products. Substituting \eqref{eq:weights}--\eqref{eq:loads}
gives the entries of $Q$ and shows that the regular price increment has zero
conditional mean. Indeed, we have
\begin{align*}
 \frac1h\sum_{i\in\sig}r_i g_i^2
 &=\frac{\sum_{i\in\sig}\omega_i\gamma_i^2}{\ell_0D_0^2}=c,\qquad
 \frac1h\sum_{i\in\sig}r_i y_i^2
 =\frac{\sum_{i\in\sig}\omega_i p_i^2}{H}=1,\\
 \frac1h\sum_{i\in\sig}r_i g_i y_i
 &=\sqrt c\,\frac{\sum_{i\in\sig}\omega_i\gamma_i p_i}
                  {\sqrt{H\sum_{i\in\sig}\omega_i\gamma_i^2}}
 =-\sqrt c\,\frac{(\beta-1)\sqrt q}{\sqrt{(1+q)(1+\beta^2q)}},\\
 \sum_{i\in\sig}r_i y_i&=\sqrt{\frac{h\ell}{H}}
                                \sum_{i\in\sig}\omega_i p_i=0.
\end{align*}
For fixed $\beta>1$, write $f_\beta(q)=\rho^2$ as in \eqref{eq:cov}. Then
\[
 f_\beta'(q)=
 \frac{(\beta-1)^2(1-\beta^2q^2)}{(1+q)^2(1+\beta^2q)^2},
 \qquad f_\beta(0)=0,\qquad
 f_\beta(1/\beta)=\left(\frac{\beta-1}{\beta+1}\right)^2 .
\]
For any prescribed $\rho_\star\in(-1,0)$, choose
$\beta>(1+|\rho_\star|)/(1-|\rho_\star|)$. Continuity and strict monotonicity
on $(0,1/\beta)$ give a unique $q$ in this interval such that
$f_\beta(q)=\rho_\star^2$. The negative sign in \eqref{eq:cov} gives $\rho=\rho_\star$.
Finally, \eqref{eq:weights}--\eqref{eq:loads} and \eqref{eq:offspring-weights} give
\[
 \mu_h(\spc)=hm_0,\qquad
 r_++r_-=2\ell,\qquad g_i=\frac{\gamma_i}{\ell_0D_0}h^\alpha .
\]
Substitution into the two ratios in \eqref{eq:rare-mark-scaling} gives
the claimed scaling.
\end{proof}

\subsection{Proof of Theorem~\ref{thm:limit}}\label{app:identification}

Before we proceed to the proof of Theorem~\ref{thm:limit}, 
we state and prove a technical lemma.
For $t\ge0$, the renewal kernel of the grid model is
\[
 k_h(t):=h^{\alpha-1}u_{\lfloor t/h\rfloor+1}.
\]
We use its $L^2$ convergence to pass to the limit in the variance convolution
and a uniform second-moment bound on $V_n$ to control the activity clock.

\begin{lemma}\label{lem:renew}
For the lag law \eqref{eq:tail}, $u_n\sim n^{\alpha-1}/\Gamma(\alpha)$ as $n\to\infty$.
As $h=T/N\downarrow0$,
\begin{equation}\label{eq:renewal-limits}
 \|k_h-\K\|_{L^2(0,T)}\longrightarrow0,\qquad
 \max_{0\le n\le N}\left|h^\alpha\sum_{j=1}^n u_j-\G(t_n)\right|\longrightarrow0.
\end{equation}
The variance satisfies \eqref{eq:mild} and, with the supremum norm on $[0,T)$,
\[
 \left\|G_h-G_0\right\|_\infty\longrightarrow0,\qquad
 \sup_{N\ge1}\max_{0\le n<N}\E\left[V_n^2\right]<\infty.
\]
\end{lemma}

\begin{proof}
The tail of the lag law is $\sum_{k=n+1}^{\infty}\phi_k=s_n$ by
\eqref{eq:tail}. It has span one since $\phi_1,\phi_2>0$.
For $n\ge1$, the renewal mass is $u_n=\sum_{j=1}^n(\phi^{*j})_n$.
Since $\alpha>1/2$, the arithmetic strong renewal theorem
\citep[Theorem~1.4]{caravenna-doney} gives
\[
 u_n\sim \frac{\sin(\pi\alpha)}{\pi n s_n}
 =\frac{\sin(\pi\alpha)\Gamma(1-\alpha)}{\pi}\,n^{\alpha-1}
 =\frac{n^{\alpha-1}}{\Gamma(\alpha)} ,
\]
where the last equality is Euler's reflection formula. Enlarging the constant
to cover finitely many indices gives $u_n\le Cn^{\alpha-1}$ for every $n\ge1$.
For $0<t\le T$, $k_h(t)\to\K(t)$ and $k_h(t)\le Ct^{\alpha-1}$.
Since $2\alpha-2>-1$, the dominated convergence theorem gives convergence
in $L^2(0,T)$. By the definition of $k_h$,
\begin{align*}
 \max_{0\le n\le N}\left|h^\alpha\sum_{j=1}^n u_j-\G(t_n)\right|
 &=\max_{0\le n\le N}\left|\int_0^{t_n}(k_h(s)-\K(s))\dd s\right|\le\sqrt T\,\|k_h-\K\|_2\longrightarrow0.
\end{align*}

Substitute \eqref{eq:drivers} into \eqref{eq:model} and divide by $a$. With
$\delta_0$ the unit sequence at zero and $U=\delta_0+u$, this gives
\[
 (I-\phi*)V=v_0s+\eta h^\alpha{\bf1}-d h^\alpha V
       +\phi*\left(\xi m_1h^\alpha V+h^{\alpha-1}\Delta M\right).
\]
The renewal recursion gives $U*(\delta_0-\phi)=\delta_0$ and $U*\phi=u$.
The tail identity $s=(\delta_0-\phi)*{\bf1}$ then gives $U*s={\bf1}$.
Convolving with $U$ and using $\xi m_1-d=b$ gives
\[
 (1+d h^\alpha)V=v_0{\bf1}+\eta h^\alpha U*{\bf1}
      +u*\left(b h^\alpha V+h^{\alpha-1}\Delta M\right),
\]
which is \eqref{eq:mild}.
For $t_n\le t<t_{n+1}$, the definition of $G_h$ gives
\[
 G_h(t)-G_0(t)
 =(a-1)G_0(t)+a\eta\left(h^\alpha+h^\alpha\sum_{j=1}^n u_j-\G(t_n)
                         +\G(t_n)-\G(t)\right).
\]
Since $a\to1$ and $\G$ is uniformly continuous on $[0,T]$,
\eqref{eq:renewal-limits} gives $\|G_h-G_0\|_\infty\to0$.

By the conditional Poisson laws and \eqref{eq:drivers}, the conditional moments satisfy
\begin{equation}\label{eq:innovation-moments}
 \E[\Delta M_j\mid\F_j]=0,\qquad
 \E\left[(\Delta M_j)^2\mid\F_j\right]=\left(c+\xi^2m_2\right)hV_j.
\end{equation}
All bin moments of order two exist by finite induction in \eqref{eq:model}.
Since $b<0$ and $V_n\ge0$,
\[
 \E\left[V_n\right]\le G_h(t_n),\qquad
 0\le V_n\le G_h(t_n)
       +a\sum_{j=0}^{n-1}h^{\alpha-1}u_{n-j}\Delta M_j.
\]
Squaring the nonnegative upper bound for $V_n$, taking expectations and using
orthogonality of the martingale differences gives
\begin{align}
 \E\left[V_n^2\right]&\le (G_h(t_n))^2+a^2\left(c+\xi^2m_2\right)
 \sum_{j=0}^{n-1}h\left(h^{\alpha-1}u_{n-j}\right)^2\E[V_j]\nonumber\\
 &\le \|G_h\|_\infty^2+\left(c+\xi^2m_2\right)\|G_h\|_\infty
                         \int_0^{t_n}\left(k_h(s)\right)^2\dd s .\label{eq:bin-second}
\end{align}
The convergence $\|G_h-G_0\|_\infty\to0$ and the $L^2(0,T)$ convergence
in \eqref{eq:renewal-limits} bound both norms for all sufficiently large $N$.
On each of the finitely many remaining grids, $G_h$ and $k_h$ are bounded
step functions on $[0,T)$. Thus, with $h=T/N$ for each positive integer $N$,
\[
 \sup_{N\ge1}\|G_h\|_\infty<\infty,\qquad
 \sup_{N\ge1}\|k_h\|_{L^2(0,T)}<\infty .
\]
Taking the maximum over $0\le n<N$ and then the supremum over $N\ge1$
in \eqref{eq:bin-second} proves the uniform second-moment bound.
\end{proof}

\subsubsection{Proof of Theorem~\ref{thm:limit}}\label{app:joint-limit-proof}

\begin{proof}
The clock $A^h$ determines both the predictable brackets of the regular
martingales and the compensator of the marked counts.

\textbf{Step 1. The activity clock and the regular order flow.}
A family in $D([0,T],\R^d)$ is \emph{$C$-tight} if it is
tight for $J_1$ and every subsequential weak limit is supported on continuous
paths. We use the piecewise constant filtration
$\mathcal F^h_t=\mathcal F_{\lfloor t/h\rfloor}$, with $\mathcal F_N$ at $T$.
We take the discrete predictable brackets in this filtration.
Let $\|\cdot\|_\infty$ denote the supremum norm on $[0,T]$, and define
\begin{equation}\label{eq:interpolated-clock}
 \bar A^h_t:=\int_0^t V^h(s)\dd s,\qquad 0\le t\le T .
\end{equation}
By the step-function definition of $V^h$ and the second-moment bound in
Lemma~\ref{lem:renew}, with $h=T/N$,
\[
 H_T:=\sup_{N\ge1}\E\left[\left\|V^h\right\|_{L^2(0,T)}^2\right]
 =\sup_{N\ge1}\sum_{n=0}^{N-1}h\E[V_n^2]
 \le T\sup_{N\ge1}\max_{0\le n<N}\E[V_n^2]<\infty .
\]
For $0\le s\le t\le T$, apply the Cauchy--Schwarz inequality to
$\bar A^h_t-\bar A^h_s=\int_s^t V^h(u)\dd u$ in \eqref{eq:interpolated-clock}.
Using $\Delta A_n=hV_n$ from \eqref{eq:drivers} gives
\begin{equation}\label{eq:clockmod}
\begin{gathered}
 \left|\bar A^h_t-\bar A^h_s\right|\le |t-s|^{1/2}\left\|V^h\right\|_2,\\
 A^h_t=\bar A^h_{h\lfloor t/h\rfloor},\qquad
 \E\left[\left\|A^h-\bar A^h\right\|_\infty^2\right]\le hH_T.
\end{gathered}
\end{equation}
On $\{\|V^h\|_2\le R\}$ the interpolated clock starts at zero and has
$1/2$-Hölder seminorm at most $R$. These paths lie in a compact subset
of $C[0,T]$ by the Arzel\`a--Ascoli theorem, and the complement has probability
at most $H_T/R^2$. Thus \eqref{eq:clockmod} proves $C$-tightness of $A^h$.
By \eqref{eq:regular-scaled},
\[
 L^{h,i}-A^h=\sqrt{\frac h{r_i}}\,Z^{h,i}.
\]
Given $\F_n$, the regular counts $N_{n,+}$ and $N_{n,-}$ are
independent Poisson variables with means $r_+V_n$ and $r_-V_n$.
The increments of \eqref{eq:regular-scaled} therefore satisfy
\[
 \operatorname{Cov}\!\left(\sqrt{h/r_i}\,\widehat N_{n,i},
           \sqrt{h/r_j}\,\widehat N_{n,j}\mid\F_n\right)
 =\ind_{\{i=j\}}hV_n,\qquad i,j\in\sig .
\]
Summing over bins and using $\Delta A_n=hV_n$ in \eqref{eq:drivers} gives
\[
 \left\langle Z^{h,i},Z^{h,j}\right\rangle_t=\ind_{\{i=j\}}A^h_t.
\]
Doob's $L^2$ maximal inequality gives
\begin{equation}\label{eq:ct-count}
 \E\left[\left\|L^{h,i}-A^h\right\|_\infty^2\right]
 \le 4(h/r_i)\E\left[A^h_T\right]\le C_T h^{2\alpha}.
\end{equation}
Here $h/r_i=h^{2\alpha}/(\omega_i\ell_0)$ by \eqref{eq:weights}.

For bounded stopping times $0\le\sigma\le\tau\le(\sigma+\delta)\wedge T$,
the two grid points in \eqref{eq:clockmod} are at most $\delta+h$ apart.
Optional sampling theorem for $Z^{h,i}$ and $(Z^{h,i})^2-A^h$ gives
\begin{equation}\label{eq:clocktight}
 \E\left[\left|{\bf Z}^h_\tau-{\bf Z}^h_\sigma\right|^2\right]
 =2\E\left[\left(A^h_\tau-A^h_\sigma\right)\right]
 \le 2\sqrt{H_T(\delta+h)} .
\end{equation}
Also,
\begin{equation}\label{eq:regular-max}
 \E\left[\sup_{t\le T}\left|{\bf Z}^h_t\right|^2\right]\le8\E\left[A^h_T\right]\le8\sqrt{TH_T}.
\end{equation}
For $R,\varepsilon>0$, Markov's inequality and \eqref{eq:clocktight} give
\begin{align}\label{two:upper:bounds}
 \Prob\!\left(\sup_{0\le t\le T}\left|{\bf Z}^h_t\right|>R\right)
 \le\frac{8\sqrt{TH_T}}{R^2},\qquad
 \sup_{\sigma,\tau}\Prob\!\left(\left|{\bf Z}^h_\tau-{\bf Z}^h_\sigma\right|>\varepsilon\right)
 \le\frac{2\sqrt{H_T(\delta+h)}}{\varepsilon^2},
\end{align}
where the supremum is over the stopping times in \eqref{eq:clocktight}. The first
bound in \eqref{two:upper:bounds} gives compact containment. Taking $\limsup_{h\downarrow0}$ and then
$\delta\downarrow0$ in the second bound in \eqref{two:upper:bounds} verifies Aldous' tightness criterion
\citep[Theorem~16.10]{billingsley} for ${\bf Z}^h=\left(Z^{h,+},Z^{h,-}\right)^\top$
in $D([0,T],\R^2)$.

Write $D_{n,i}:=\sqrt{h/r_i}\,\widehat N_{n,i}$. A Poisson variable of mean
$\lambda$ has fourth centered moment $\lambda+3\lambda^2$. Hence
\begin{equation}\label{eq:ct-jumps}
\begin{aligned}
 \E\left[D_{n,i}^4\mid\F_n\right]&=\frac{h^2}{r_i}V_n+3h^2V_n^2,\\
 \sum_{n=0}^{N-1}\E\left[D_{n,i}^4\right]&\le C_T\left(h^{2\alpha}+h\right),\\
 \Prob\!\left(\max_{n<N}|\Delta{\bf Z}_n|>\varepsilon\right)
 &\le C_T\varepsilon^{-4}\left(h^{2\alpha}+h\right).
\end{aligned}
\end{equation}
Indeed, $|\Delta{\bf Z}_n|^4=(D_{n,+}^2+D_{n,-}^2)^2
\le2(D_{n,+}^4+D_{n,-}^4)$. Thus the union bound and Markov's inequality give
\[
 \Prob\!\left(\max_{0\le n<N}|\Delta{\bf Z}_n|>\varepsilon\right)
 \le \varepsilon^{-4}\sum_{n=0}^{N-1}\E\left[|\Delta{\bf Z}_n|^4\right]
 \le 2\varepsilon^{-4}\sum_{i\in\sig}\sum_{n=0}^{N-1}\E\left[D_{n,i}^4\right],
\]
which gives the last bound in \eqref{eq:ct-jumps}.
If $x_k\to x$ in vector $J_1$, there are time changes $\lambda_k$
with $\|x_k\circ\lambda_k-x\|_\infty\to0$. Taking left limits gives
\[
 \sup_{0<t\le T}|\Delta x(t)|
 \le \sup_{0<t\le T}|\Delta x_k(t)|+2\left\|x_k\circ\lambda_k-x\right\|_\infty .
\]
Thus, \eqref{eq:ct-jumps} makes every limit of the tight martingale vector
continuous. The clock and the martingale vector are jointly $C$-tight since
all their marginal limits are continuous and $J_1$ convergence to continuous
paths gives uniform convergence. Equation~\eqref{eq:ct-count} transfers this
to the count vector. By \eqref{eq:regular-map}, adjoining the two regular drivers $B^h,W^h$
preserves joint $C$-tightness.

For $i,j\in\sig$, the error in the optional covariation is the martingale
\[
 U^{h,ij}_t:=\left[Z^{h,i},Z^{h,j}\right]_t-\ind_{\{i=j\}}A^h_t
 =\sum_{\substack{0\le n<N\\t_{n+1}\le t}}
       \left(D_{n,i}D_{n,j}-\ind_{\{i=j\}}hV_n\right).
\]
Its conditional increment variance is
\[
 \operatorname{Var}\left(D_{n,i}D_{n,j}\mid\F_n\right)=
 \begin{cases}
 (h^2/r_i)V_n+2h^2V_n^2,&i=j,\\
 h^2V_n^2,&i\ne j.
 \end{cases}
\]
The diagonal follows from \eqref{eq:ct-jumps}, and the off-diagonal follows
by conditional independence. Summing these variances and applying Doob's
$L^2$ maximal inequality yields
\begin{equation}\label{eq:ct-bracket}
 \max_{i,j\in\sig}\E\left[\left\|U^{h,ij}\right\|_\infty^2\right]\le C_T\left(h^{2\alpha}+h\right)\longrightarrow0.
\end{equation}

\textbf{Step 2. Tightness of the marked coordinates.}
Let $Y^h_\circ:=(B^h,W^h,C^h,J^h)$ as in \eqref{eq:drivers}.
Its third coordinate $C^h$ counts marks and its fourth coordinate $J^h$
sums their positive sizes, so both are increasing. The processes $C^h-m_0A^h$ and $J^h-m_1A^h$
are square-integrable martingales by conditional Poisson sampling and $m_2<\infty$.
For the stopping times of \eqref{eq:clocktight}, optional sampling theorem gives
\begin{align*}
 \E\left[\left(C^h_\tau-C^h_\sigma\right)+\left(J^h_\tau-J^h_\sigma\right)\right]
 &=(m_0+m_1)\E\left[\left(A^h_\tau-A^h_\sigma\right)\right]\le(m_0+m_1)\sqrt{H_T(\delta+h)}.
\end{align*}
Also,
\[
 \E\left[\left(C^h_T+J^h_T\right)\right]=(m_0+m_1)\E\left[A^h_T\right]\le(m_0+m_1)\sqrt{TH_T}.
\]
Markov's inequality and Aldous' criterion
\citep[Theorem~16.10]{billingsley} give tightness of the \emph{pair} $(C^h,J^h)$
in vector $J_1$. Adjoining the $C$-tight regular coordinates and clock from Step~1
preserves vector $J_1$ tightness because, along a convergent product subsequence,
the time changes for $(C^h,J^h)$ also work for the uniformly converging continuous-limit
coordinates. The conditional factorial moment of the Poisson count gives
\begin{equation}\label{eq:double}
\begin{aligned}
 \Prob(\exists n<N:\Delta C_n\ge2)
 &\le\frac12\sum_{n=0}^{N-1}\E[\Delta C_n(\Delta C_n-1)]=\frac12m_0^2h^2\sum_{n=0}^{N-1}\E\left[V_n^2\right]
 \le C_T h.
\end{aligned}
\end{equation}
Fix a deterministic $t\in[0,T]$ and a radius $\delta>0$.
Let 
\[
I_\delta(t):=[t-\delta,t+\delta]\cap[0,T] 
\quad\text{and}\quad
E_h(t,\delta):=\left\{0\le n<N:t_{n+1}\in I_\delta(t)\right\}.
\]
There are at most $2\delta/h+2$ such endpoints. Since
$\E[\Pi_n(\spc)\mid\F_n]=hm_0V_n$,
\begin{align}
 \Prob\left(\J^h(I_\delta(t)\times\spc)>0\right)
 &\le\sum_{n\in E_h(t,\delta)}\E[\Pi_n(\spc)]
 \le m_0(2\delta+2h)\max_{0\le n<N}\E[V_n].\label{eq:mark-window}
\end{align}
Along a subsequence converging in vector $J_1$, take a Skorokhod representation
of $(C^h,J^h)$ with limit $(C,J)$ and time changes $\lambda_h$ for which
\[
\left\|\lambda_h-\mathrm{id}\right\|_\infty+
\left\|\left(C^h,J^h\right)\circ\lambda_h-(C,J)\right\|_\infty\to0,
\]
almost surely.
For $t\in(0,T]$, the aligned jump converges to $(\Delta C_t,\Delta J_t)$,
since taking left limits preserves the uniform bound. A nonzero limit jump
therefore requires a marked endpoint in $I_\delta(t)$ eventually.
Along this subsequence, Fatou's lemma and \eqref{eq:mark-window} give the first
two inequalities below. The Cauchy--Schwarz inequality gives the last one.
\begin{align*}
 \Prob\left(\left(\Delta C_t,\Delta J_t\right)\ne(0,0)\right)
 &\le \liminf_{h\downarrow0}\Prob\left(\J^h(I_\delta(t)\times\spc)>0\right)\\
 &\le 2m_0\delta\sup_{N\ge1}\max_{0\le n<N}\E[V_n]\\
 &\le 2m_0\delta\left(\sup_{N\ge1}\max_{0\le n<N}\E[V_n^2]\right)^{1/2}.
\end{align*}
In these suprema $h=T/N$. By Lemma~\ref{lem:renew}, the final bound tends to zero
as $\delta\downarrow0$, so $(C,J)$ has no jump at this fixed $t$ almost surely.
At $t=0$, $C^h_0=J^h_0=0$, and
\[
\lim_{\delta\downarrow0}\limsup_{h\downarrow0}
\Prob\left(\J^h([0,\delta]\times\spc)>0\right)=0,
\]
by \eqref{eq:mark-window} with $t=0$.
Thus no mark can accumulate at the initial endpoint.

\textbf{Step 3. The singular convolution.}
Let $D^h_\circ=bA^h+M^h$, where \eqref{eq:drivers} gives
\[
 M^h_t=\sum_{\substack{0\le j<N\\t_{j+1}\le t}}
          \left[\Delta B_j+\xi\left(Z_j-m_1hV_j\right)\right]
       = B^h_t+\xi\left(J^h_t-m_1 A^h_t\right).
\]
Step~2 gives vector $J_1$ tightness of $\left(A^h,B^h,W^h,C^h,J^h\right)$.
The same time change works for its linear image
\[
D^h_\circ=\left(b-\xi m_1\right)A^h+B^h+\xi J^h,
\]
so that $D^h_\circ$ is jointly tight
with these coordinates.
For $f\in C_c^\infty((0,\infty))$, extended by zero, define
\[
\mathcal T_f(x)(t):=\int_0^t f'(t-s)x(s)\dd s\qquad\text{on $D([0,T],\R)$.}
\]
Integration by parts for a càdlàg semimartingale $D$ with $D_0=0$ gives
\begin{equation}\label{eq:smooth-convolution-representation}
 \int_{(0,t]} f(t-s)\dd D_s
 =f(0)D_t-f(t)D_0+\int_0^t f'(t-s)D_s\dd s=\mathcal T_f(D)(t),
\end{equation}
since $f(0)=0$. If $x_j\to x$ in $J_1$, the paths are uniformly bounded
and converge at every continuity point of $x$. Thus,
\[
\left\|\mathcal T_f(x_j)-\mathcal T_f(x)\right\|_\infty
\le\|f'\|_\infty\left\|x_j-x\right\|_{L^1(0,T)}\to0.
\]
For deterministic lags $l\in L^2(0,T)$, Fubini's theorem, orthogonality of the martingale differences in
\eqref{eq:innovation-moments} and the Cauchy--Schwarz inequality for the drift sum give
\begin{align*}
 &\E\left[\int_0^T\left|\sum_{0\le j<N,t_{j+1}\le t}
       l(t-t_{j+1})\Delta M_j\right|^2\dd t\right]\le\left(c+\xi^2m_2\right)\sum_{j=0}^{N-1}h\E[V_j]\,\|l\|_2^2,\\
 &\E\left[\int_0^T\left|\sum_{0\le j<N,t_{j+1}\le t}
       l(t-t_{j+1})hV_j\right|^2\dd t\right]\le T\sum_{j=0}^{N-1}h\E\left[V_j^2\right]\,\|l\|_2^2.
\end{align*}
For $t_n\le t<t_{n+1}$ and $0\le j<n$,
$\lfloor(t-t_{j+1})/h\rfloor+1=n-j$.
Thus the coefficient $h^{\alpha-1}u_{n-j}$ in \eqref{eq:mild} is precisely
$k_h(t-t_{j+1})$. Applying the two bounds with $l=k_h-f$ gives
\begin{equation}\label{eq:smooth}
 \E\left[\left\|V^h-G_h-a\mathcal T_f\left(D^h_\circ\right)\right\|_2^2\right]
 \le C_T\|k_h-f\|_2^2.
\end{equation}
Choose $f_p\in C_c^\infty((0,T))$ with
\begin{equation}\label{eq:smooth-kernel-approximation}
 \|f_p-\K\|_2\xrightarrow[p\to\infty]{}0.
\end{equation}
Let $V^{h,p}:=G_h+a\mathcal T_{f_p}(D^h_\circ)$.
From any sequence $h\downarrow0$, extract a subsequence for which the vector
$\mathcal Y^h=\left(A^h,B^h,W^h,C^h,J^h,D^h_\circ\right)$ converges to $\mathcal Y$
in vector $J_1$. Write $D_\circ$ for its last coordinate. For each fixed
$p,q$, the continuous mapping theorem, $a\to1$ and $\|G_h-G_0\|_\infty\to0$ give
\[
 \left(\mathcal Y^h,V^{h,p},V^{h,q}\right)\ \Longrightarrow\
 \left(\mathcal Y,V^{p},V^{q}\right),\qquad
 V^{p}=G_0+\mathcal T_{f_p}(D_\circ).
\]
By lower semicontinuity of the squared norm, \eqref{eq:smooth} and
$\|k_h-\K\|_2\to0$,
\[
 \E\left[\left\|V^p-V^q\right\|_2^2\right]
 \le\liminf_{h\downarrow0}\E\left[\left\|V^{h,p}-V^{h,q}\right\|_2^2\right]
 \le 2C_T\left(\|\K-f_p\|_2^2+\|\K-f_q\|_2^2\right).
\]
Completeness of $L^2(\Omega;L^2(0,T))$ gives $V^p\to V$ in that space
(the smooth convolutions are square integrable by the bounds in
Lemma~\ref{lem:renew} and \eqref{eq:smooth}).
For every $\varepsilon>0$, Markov's inequality gives
\[
 \limsup_{h\downarrow0}\Prob\left(\left\|V^h-V^{h,p}\right\|_2>\varepsilon\right)
 \le C_T\varepsilon^{-2}\left\|\K-f_p\right\|_2^2\xrightarrow[p\to\infty]{}0.
\]
The converging-together theorem \citep[Theorem~3.2]{billingsley}, applied to
the pairs $\left(\mathcal Y^h,V^{h,p}\right)$, yields
$\left(\mathcal Y^h,V^h\right)\Rightarrow(\mathcal Y,V)$ along this subsequence, with
the norm topology on $L^2(0,T)$.
By \eqref{eq:clockmod},
\[
 \left\|A^h-\int_0^\cdot V^h(s)\dd s\right\|_\infty
 \longrightarrow0\qquad\text{in probability}.
\]
The integration
map is continuous from $L^2(0,T)$ to $C[0,T]$ by the Cauchy--Schwarz inequality, so
$A_t=\int_0^t V_s\dd s$ simultaneously for $0\le t\le T$ almost surely.
The positive cone in $L^2$ is closed, so $V\ge0$ almost everywhere almost surely.
The uniform count approximation in \eqref{eq:ct-count} gives $L^+=L^-=A$.
Equation~\eqref{eq:regular-map} also gives
\[
 {\bf Z}=\mathsf P^{-1}(B,W)^\top.
\]

\textbf{Step 4. Characteristics in the joint filtration.}
For $\theta=(\theta_1,\ldots,\theta_4)\in\R^4$, write
\[
 \theta_r:=(\theta_1,\theta_2),\qquad l_i:=\theta_1g_i+\theta_2y_i.
\]
The conditional characteristic function of one
increment of $Y^h_\circ$ is $\exp\{hV_n P_h(\theta)\}$, where
\begin{align}
 P_h(\theta)&=h^{-1}\sum_{i\in\sig}r_i\left(\e^{\ii l_i}-1-\ii l_i\right)+\int_{\spc}\left(\e^{\ii(\theta_3+\theta_4 z)}-1\right)\nu(\dd z),\nonumber\\
 P(\theta)&=-\frac12\theta_r Q\theta_r^\top
        +\int_{\spc}\left(\e^{\ii(\theta_3+\theta_4 z)}-1\right)\nu(\dd z).
 \label{eq:symbol}
\end{align}
The Taylor remainder and \eqref{eq:cov} imply
\[
 |P_h(\theta)-P(\theta)|\le C_\theta h^\alpha.
\]
For fixed $\theta$, $\mathrm{Re}\,P_h\le0$. The integral Taylor formula gives
$|\e^z-1-z|\le |z|^2/2$ for $\mathrm{Re}\,z\le0$. Thus Lemma~\ref{lem:renew}
and \eqref{eq:symbol} give
\begin{align}\label{eq:symbol-remainder}
\E\left[\sum_{n=0}^{N-1}
   \left|\e^{hV_nP_h(\theta)}-1-hV_nP(\theta)\right|\right]&\le \frac12|P_h(\theta)|^2h^2\sum_{n=0}^{N-1}\E\left[V_n^2\right]
       +|P_h(\theta)-P(\theta)|\E\left[A^h_T\right]\nonumber\\
 &\le C_\theta\left(h+h^\alpha\right)\longrightarrow0.
\end{align}
It follows for each subsequential limit $Y_\circ=(B,W,C,J)$ that
\begin{equation}\label{eq:test}
 \e^{\ii\theta\cdot Y_{\circ,t}}-
   P(\theta)\int_0^t \e^{\ii\theta\cdot Y_{\circ,s-}}\dd A_s
 \quad\text{is a martingale}.
\end{equation}
Fix $0\le s<t\le T$. Let $F_s^h$ be a bounded continuous function of
$V^h|_{(0,s)}$ in $L^2(0,s)$ and of finitely many values of
$\left(A^h,Y_\circ^h\right)$ at deterministic times $r\le s$. These variables are
$\mathcal F_s^h$-measurable. Write $F_s$ for the same function of the limit
coordinates. Step~2 excludes fixed jumps, so on a joint Skorokhod representation
$F_s^h\to F_s$ almost surely. With $E^h_r=\exp(\ii\theta\cdot Y^h_{\circ,r})$,
conditional expectation bin by bin gives
\begin{equation}\label{eq:grid-test-identity}
 \E\!\left[F_s^h\!\left(E^h_t-E^h_s-
 \sum_{0\le n<N,s<t_{n+1}\le t}
 E^h_{t_n}\left\{\exp\left(hV_nP_h(\theta)\right)-1\right\}\right)\right]=0.
\end{equation}
By \eqref{eq:symbol-remainder}, replacing
$\exp(hV_nP_h(\theta))-1$ by $hV_nP(\theta)$ in
\eqref{eq:grid-test-identity} gives an $L^1$ error tending to zero. Define
$u_h:=h\lfloor u/h\rfloor$ and $E_r:=\exp(\ii\theta\cdot Y_{\circ,r})$.
Since $E^h_r=E^h_{t_n}$ and $V^h(r)=V_n$ for $t_n\le r<t_{n+1}$,
the sum in \eqref{eq:grid-test-identity} after this replacement satisfies
\begin{equation}\label{eq:grid-test-integral}
 \sum_{\substack{0\le n<N\\s<t_{n+1}\le t}}E^h_{t_n}hV_nP(\theta)
 =P(\theta)\sum_{n=\lfloor s/h\rfloor}^{\lfloor t/h\rfloor-1}
       \int_{t_n}^{t_{n+1}}E^h_rV^h(r)\dd r=P(\theta)\int_{s_h}^{t_h}E^h_rV^h(r)\dd r,
\end{equation}
with an empty sum equal to zero. The integral in
\eqref{eq:grid-test-integral} satisfies
\begin{equation}\label{eq:grid-test-integral-error}
\begin{aligned}
 \left|\int_{s_h}^{t_h}E^h_rV^h(r)\dd r-\int_s^t E_{r-}V_r\dd r\right|
 &\le 2\sqrt h\,\left\|V^h\right\|_2+\sqrt T\left\|V^h-V\right\|_2\\
 &\quad+\int_s^t \left|E^h_r-E_{r-}\right|V_r\dd r .
\end{aligned}
\end{equation}
The first two terms on the right-hand side of
\eqref{eq:grid-test-integral-error} tend to zero almost surely on the
representation. For the last term, $E^h_r\to E_{r-}$ at every continuity
point of $Y_\circ$, and $2V\in L^1(0,T)$, so the dominated convergence
theorem applies. The integrals in \eqref{eq:grid-test-integral} are uniformly integrable since their
absolute values are bounded by $\sqrt T\|V^h\|_2$ and
$\sup_h\E\|V^h\|_2^2<\infty$ by Lemma~\ref{lem:renew}.
Passing to the limit in \eqref{eq:grid-test-identity} using
\eqref{eq:symbol-remainder}, \eqref{eq:grid-test-integral} and
\eqref{eq:grid-test-integral-error} gives
\begin{equation}\label{eq:limit-test-identity}
 \E\!\left[F_s\!\left(E_t-E_s-P(\theta)\int_s^t E_{r-}\dd A_r\right)\right]=0 .
\end{equation}
The monotone class theorem extends \eqref{eq:limit-test-identity} to the joint past generated by
$V|_{(0,s)}$ and $(A,Y_\circ)|_{[0,s]}$.
For $s_j\downarrow s$, the process in \eqref{eq:test} converges at $s_j$
to its value at $s$ by right continuity, and its absolute value is bounded
by $1+|P(\theta)|A_T$. The dominated convergence theorem gives \eqref{eq:limit-test-identity} in the
right-continuous augmentation. Likewise,
\[
\E\left[F_s^h\left(\left(B^h_t,W^h_t\right)-\left(B^h_s,W^h_s\right)\right)\right]=0. 
\]
The uniform $L^2$ maximal
bound \eqref{eq:regular-max} and the linear relation \eqref{eq:regular-map}
give uniform integrability of these increments. Passing to the limit proves
that $(B,W)$ is a continuous martingale in the same filtration.

The other coordinates $C,J$ are increasing, so $Y_\circ=(B,W,C,J)$ is a
semimartingale. Let $(\mathcal B,\mathcal C,\mathcal N)$ be its predictable
characteristics for $\chi(y)=y\ind_{\{|y|\le1\}}$, and let
$\iota(z):=(0,0,1,z)$, $Q_4:=\operatorname{diag}(Q,0_{2\times2})$.
Applying It\^o's formula to $E_t=\exp(\ii\theta\cdot Y_{\circ,t})$
and comparing its predictable finite-variation part with
$P(\theta)\int_0^t E_{s-}\dd A_s$ in \eqref{eq:test} gives,
after division by $E_{t-}\ne0$,
\begin{equation}\label{eq:characteristic-identity}
 \ii\theta\cdot\dd\mathcal B_t-\frac12\theta\,\dd\mathcal C_t\,\theta^\top
 +\int_{\R^4}\left(\e^{\ii\theta\cdot y}-1-\ii\theta\cdot\chi(y)\right)
                   \mathcal N(\dd t,\dd y)
 =P(\theta)\dd A_t.
\end{equation}
Take Radon--Nikodym densities of both sides of
\eqref{eq:characteristic-identity} with respect to the locally finite
dominating measure
\[
 \dd A_t+|\dd\mathcal B_t|+\operatorname{tr}(\dd\mathcal C_t)
 +\int_{\R^4}\left(1\wedge|y|^2\right)\mathcal N(\dd t,\dd y).
\]
The density equality obtained from \eqref{eq:characteristic-identity}
holds outside a common null set for all $\theta\in\mathbb Q^4$, and
continuity in $\theta$ extends it to every $\theta\in\R^4$.
Since $|\iota(z)|>1$ for $z>0$,
$\chi(\iota(z))=0$. Uniqueness of the L\'evy--Khintchine triplet applied to
\eqref{eq:characteristic-identity} with the symbol \eqref{eq:symbol} gives
\[
 \dd\mathcal B_t=0,\qquad \dd\mathcal C_t=Q_4\,\dd A_t,\qquad
 \mathcal N(\dd t,\dd y)=\dd A_t\int_{\spc}\delta_{\iota(z)}(\dd y)\nu(\dd z).
\]
The expected number of jumps outside $\iota(\spc)$ is zero, and the expected
number in $\iota(\spc)$ is $m_0\E [A_T]<\infty$.
The semimartingale decomposition for $\chi$ now gives
\[
 (C_t,J_t)=\sum_{\substack{0<r\le t\\\Delta C_r=1}}(1,\Delta J_r),\qquad
 \J=\sum_{\substack{0<r\le T\\\Delta C_r=1}}\delta_{(r,\Delta J_r)}.
\]
Thus $\J$ is simple in time. The continuous bracket matrix and the
predictable compensator of $\J$ are
\begin{equation}\label{eq:limit-brackets-compensator}
 \left\langle(B,W)^\top\right\rangle_t=Q A_t,\qquad
 \nu^{\J}(\dd t,\dd z)=\dd A_t\,\nu(\dd z).
\end{equation}
Using \eqref{eq:limit-brackets-compensator} in \eqref{eq:regular-map} gives
\[
\langle{\bf Z}\rangle=\mathsf P^{-1}Q\left(\mathsf P^{-1}\right)^\top A=I_2A.
\]
Every square-integrable integral against $\J-\nu^\J$ is purely discontinuous,
so that its covariations with $B,W$ vanish.

To recover convergence of the marked measures, define
$\widehat{\J}^h:=\sum_{0<r\le T:\,\Delta C^h_r=1}\delta_{(r,\Delta J^h_r)}$,
a function of $(C^h,J^h)$. By \eqref{eq:double},
$\Prob(\widehat{\J}^h\ne\J^h)\le C_T h$.
On the joint representation, the common time changes satisfy
$\|C^h\circ\lambda_h-C\|_\infty<1/2$ eventually. The grid and limiting
marked counts $C^h_t=\J^h((0,t]\times\spc)$ and $C_t=\J((0,t]\times\spc)$
are integer valued. Thus
$C^h_{\lambda_h(t)}=C_t$ for every $t\in[0,T]$ once the bound holds.
If $r_1,\ldots,r_{C_T}$ are the jump times of $C$, the jump times of $C^h$
are $\lambda_h(r_j)$, and
\[
 \left|\Delta J^h_{\lambda_h(r_j)}-\Delta J_{r_j}\right|
       \le2\left\|J^h\circ\lambda_h-J\right\|_\infty\to0.
\]
For $g\in C_c([0,T]\times\spc)$ this gives
\[
 \int_{[0,T]\times\spc} g\,\dd\widehat{\J}^h
 =\sum_{j=1}^{C_T}g\left(\lambda_h(r_j),\Delta J^h_{\lambda_h(r_j)}\right)
 \longrightarrow \sum_{j=1}^{C_T}g\left(r_j,\Delta J_{r_j}\right)
 =\int_{[0,T]\times\spc} g\,\dd\J .
\]
Since $\Prob(\widehat{\J}^h\ne\J^h)\to0$, the same joint weak convergence
holds for $\J^h$ and the coordinates in Step~3.

\textbf{Step 5. Identification and uniqueness.}
The identity $D_\circ=bA+M$ follows from the linear identity in Step~3, with
$M=B+\xi(J-m_1A)$. The bracket and compensator in
\eqref{eq:limit-brackets-compensator} give
$\langle M\rangle=(c+\xi^2m_2)A$ because $B$ is continuous and
$J-m_1A$ is a purely discontinuous martingale. Lower semicontinuity
gives $\E[\|V\|_2^2]\le H_T$. For $l_p=f_p-\K$, It\^o's isometry and
Tonelli's theorem give
\begin{align}
 \E\left[\int_0^T\left|\int_{[0,t)} l_p(t-s)\dd M_s\right|^2\dd t\right]
 &=\left(c+\xi^2m_2\right)\E\left[\int_0^T V_s
                                \int_s^T |l_p(t-s)|^2\dd t\,\dd s\right]\nonumber\\
 &\le\left(c+\xi^2m_2\right)\E[A_T]\|l_p\|_2^2 .
 \label{eq:limit-martingale-kernel-error}
\end{align}
Applying the Cauchy--Schwarz inequality to the drift integral gives
\begin{equation}\label{eq:limit-drift-kernel-error}
 \E\left[\int_0^T\left|b\int_0^t l_p(t-s)V_s\dd s\right|^2\dd t\right]
 \le b^2T\E\left[\|V\|_2^2\right]\|l_p\|_2^2.
\end{equation}
The mean-square errors from replacing $\K$ by $f_p$ in the
martingale and drift convolutions, bounded in
\eqref{eq:limit-martingale-kernel-error} and \eqref{eq:limit-drift-kernel-error},
tend to zero as $p\to\infty$ by \eqref{eq:smooth-kernel-approximation}.
By \eqref{eq:smooth-convolution-representation} with $D_\circ=bA+M$,
the smooth approximations in Step~3 satisfy
\[
 V^p(t)=G_0(t)+b\int_0^t f_p(t-s)V_s\dd s+
                  \int_{[0,t)} f_p(t-s)\dd M_s .
\]
Since $V^p\to V$ in $L^2(\Omega;L^2(0,T))$ by Step~3, the two
mean-square estimates identify $V$ with the right-hand side of
\eqref{eq:target} in that space.

For the price, $\sum_{i\in\sig}r_i y_i=0$ and
$\sum_{i\in\sig}r_i y_i^2=h$ by \eqref{eq:cov} and its proof. Hence
\eqref{eq:price} yields $X^h=W^h-\Lambda J^h-(D_h/h)A^h$, with
\[
 \left|\frac{D_h}{h}-\kappa\right|
 =\frac1h\left|\sum_{i\in\sig}r_i
          \left(\e^{y_i}-1-y_i-\frac12y_i^2\right)\right|
 \le \frac{\e^{\max_i|y_i|}}{6h}\sum_{i\in\sig}r_i|y_i|^3=\mathcal O(h^\alpha).
\]
The bound uses $\max_i|y_i|=\mathcal O(h^\alpha)$ from \eqref{eq:loads}. Thus vector
$J_1$ convergence of $(A^h,B^h,W^h,J^h)$ identifies
$X=W-\Lambda J-\kappa A$ jointly with $V,B,W,\J$.

For uniqueness, normalize the variance martingale by
$\widetilde B=B/\sqrt c$. The bracket matrix in
\eqref{eq:limit-brackets-compensator} and the entries of $Q$ in
\eqref{eq:cov} give
\[
 \left\langle\widetilde B\right\rangle=\langle W\rangle=A,\qquad
 \left\langle\widetilde B,W\right\rangle=\rho A.
\]
Both normalized drivers are continuous local martingales and are strongly
orthogonal to stochastic integrals against
$\J(\dd t,\dd z)-\dd A_t\,\nu(\dd z)$, which are purely discontinuous.
Since $A,B$ are continuous and $\J$ is simple in time,
the jump measure of $D_\circ=bA+B+\xi(J-m_1A)$ is
\[
 \sum_{\substack{0<s\le T\\\Delta D_{\circ,s}\ne0}}
      \delta_{(s,\Delta D_{\circ,s})}
   =(\mathrm{id}\times(z\mapsto\xi z))_\#\J,
\]
where $\#$ denotes pushforward. Thus
$(V,\widetilde B,W,\J,X)$ satisfies the complete canonical system of
\citet[Definition~5.4]{wwz}, with their parameters
$(\bar c_B,\bar\xi,\nu_J)=(c,\xi,\nu)$.
Indeed, the restricted $L^2(0,t)$ class of $V$ is recovered from the past
of $A$ by left differentiation. Conversely, $A_t=\int_0^t V_s\dd s$.
The paths of $C,J$ up to $t$ are determined by the marks up to $t$,
and Step~4 recovers those marks from their jumps. Thus the filtration in
Step~4 agrees with the completed, right-continuous canonical filtration.
Their Assumption~2.1 holds because $\Lambda\ge0$, $m_0,m_2<\infty$, and
$\nu$ is supported on $\spc$. The initial-curve condition of their
Assumption~2.3 holds for the continuous, nondecreasing function
$G_0(t)=v_0+\eta\G(t)$, defined for all $t\ge0$.
Finally $c,\xi>0$ and $-1<\rho<0$ by \eqref{eq:cov}.
To apply \citet[Proposition~5.5]{wwz} on its stated unit horizon, set
$\widehat V_u:=T V_{Tu}$, $\widehat A_u:=A_{Tu}$ for $0\le u\le1$, and
record each mark $(s,z)$ at $(s/T,z)$. The normalized martingales
$\widetilde B_{Tu},W_{Tu}$ have bracket matrix
$\left(\begin{smallmatrix}1&\rho\\\rho&1\end{smallmatrix}\right)\widehat A_u$,
and the mark compensator is $\dd\widehat A_u\,\nu(\dd z)$.
Since $T\K(Tu)=T^\alpha\K(u)$, multiplying \eqref{eq:target} at $Tu$
by $T$ gives the unit-horizon parameters
\[
 \widehat G_0(u)=T G_0(Tu),\qquad
 \left(\widehat b,\sqrt{\widehat c},\widehat\xi\right)
 =T^\alpha\left(b,\sqrt c,\xi\right).
\]
The log price $X_{Tu}$ satisfies \eqref{eq:tprice} with clock $\widehat A$
and the same price loading. The rescaled curve $\widehat G_0$ is continuous
and nondecreasing, with $\widehat G_0(0)=Tv_0\ge0$.
The time change preserves the sign and integrability conditions above.
\citet[Proposition~5.5]{wwz} therefore give weak uniqueness of the complete
canonical law on $[0,1]$. Inverting the time change gives
the unique complete canonical law $\mathcal P_\star$ of
$\left(V,\widetilde B,W,\J,X\right)$ on $[0,T]$. Define
$F\left(v,\widetilde b,w,\mu,x\right):=\left(v,\sqrt c\,\widetilde b,w,x,\mu\right)$. For every
subsequential law $\mathcal P$ identified above,
\[
 \mathcal P=\Law(V,B,W,X,\J)=F_\#\mathcal P_\star .
\]
Every sequence of meshes tending to zero has a subsequence with this law as its
limit, which proves convergence along the full sequence.
This completes the proof.
\end{proof}


\subsection{Proof of Corollary~\ref{cor:european-prices}}\label{app:european-prices}

\begin{proof}
If $x_j\to x$ in $J_1$ on $[0,T]$, the associated time changes fix $T$,
so $x_j(T)\to x(T)$. The continuous mapping theorem applied to
Theorem~\ref{thm:limit} therefore gives
\[
 S^h_T=S_0\e^{X^h_T}\ \Longrightarrow\ S_0\e^{X_T}=S_T.
\]
Since $s\mapsto(K-s)^+$ is continuous and bounded by $K$ on
$[0,\infty)$, the first limit in \eqref{eq:european-price-convergence}
follows.

For the call option, we verify the martingale property of $S$ using
\citet[Theorem~3.3 and Corollary~3.4]{bps}. Let
$L_\alpha(\dd s)=s^{-\alpha}\dd s/\Gamma(1-\alpha)$.
This measure has a nonnegative, nonincreasing, locally integrable density.
For $x,y>0$, Euler's beta integral defines the beta function
\citep[Eq.~(5.12.1)]{askey-roy}.
\begin{equation}\label{eq:euler-beta}
 \mathrm B(x,y):=\int_0^1 v^{x-1}(1-v)^{y-1}\dd v
 =\frac{\Gamma(x)\Gamma(y)}{\Gamma(x+y)}.
\end{equation}
With $v=s/t$ and $(x,y)=(1-\alpha,\alpha)$ in \eqref{eq:euler-beta},
we obtain
\begin{equation}\label{eq:kernel-resolvent}
 \int_0^t\K(t-s)L_\alpha(\dd s)
 =\frac{\mathrm B(1-\alpha,\alpha)}
        {\Gamma(\alpha)\Gamma(1-\alpha)}=1,
 \qquad t>0.
\end{equation}
Thus $L_\alpha$ is the resolvent of the first kind of $\K$.
For $\varepsilon>0$, the substitution $v=s/(t+\varepsilon)$ gives the
shifted convolution
\begin{equation}\label{eq:shifted-kernel-resolvent}
 q_\varepsilon(t):=\int_0^t\K(t+\varepsilon-s)L_\alpha(\dd s)
 =\frac{1}{\Gamma(\alpha)\Gamma(1-\alpha)}
   \int_0^{t/(t+\varepsilon)}v^{-\alpha}(1-v)^{\alpha-1}\dd v,
 \qquad t\ge0.
\end{equation}
The upper limit $t/(t+\varepsilon)$ in \eqref{eq:shifted-kernel-resolvent}
increases with $t$ and lies in $[0,1)$. Since the integrand is nonnegative,
$q_\varepsilon$ is nondecreasing and
\[
 0\le q_\varepsilon(t)
 \le \frac{\mathrm B(1-\alpha,\alpha)}{\Gamma(\alpha)\Gamma(1-\alpha)}=1
\]
by \eqref{eq:euler-beta}. Consequently,
\begin{equation}\label{eq:shifted-kernel-variation}
 \sup_{0<\varepsilon\le T}
 \operatorname{Var}_{[0,T]}(q_\varepsilon)
 =\sup_{0<\varepsilon\le T}\bigl(q_\varepsilon(T)-q_\varepsilon(0)\bigr)
 \le1,
\end{equation}
where $\operatorname{Var}_{[0,T]}$ denotes total variation.
The kernel $\K$ is positive, nonincreasing and continuously
differentiable on $(0,\infty)$, and $\K\in L^2(0,T)$ because
$\alpha>1/2$. Also, $G_0(t)=v_0+\eta\G(t)$ is continuous and
nondecreasing with $G_0(0)\ge0$.
Under the change of mark $y=\xi z$, the measure
$\nu_\xi:=(z\mapsto\xi z)_\#\nu$ satisfies
$\int_{\spc}y^2\nu_\xi(\dd y)=\xi^2m_2<\infty$,
and the price loading is $\Lambda/\xi\ge0$.
These properties and \eqref{eq:kernel-resolvent}--\eqref{eq:shifted-kernel-variation} verify the
hypotheses of the cited results for \eqref{eq:target}--\eqref{eq:tprice}.
Consequently $S$ is a martingale. Together with
Proposition~\ref{prop:model}, this gives
\begin{equation}\label{eq:stock-means}
 \E\left[S^h_t\right]=\E[S_t]=S_0,\qquad 0\le t\le T.
\end{equation}
Taking expectations in $(s-K)^+-(K-s)^+=s-K$ and using
\eqref{eq:stock-means}, we obtain
\begin{align*}
 \E\left[\left(S^h_T-K\right)^+\right]
 &=\E\left[\left(K-S^h_T\right)^+\right]+S_0-K
 \\
 &\longrightarrow\E\left[\left(K-S_T\right)^+\right]+S_0-K=\E\left[\left(S_T-K\right)^+\right].
\end{align*}
This proves the second limit in \eqref{eq:european-price-convergence}.
\end{proof}


\subsection{Proof of Lemma~\ref{lem:barrier-atom}}\label{app:barrier-atom-proof}

\begin{proof}
On a product extension, let $D^\perp$ be a standard Brownian motion
independent of the original system. With $A_t=\int_0^t V_s\dd s$, let
\[
 U_t:=\int_0^t\sqrt{V_s}\dd D^\perp_s,\qquad
 \sigma_\perp:=\sqrt{1-\rho^2}>0,\qquad
 W'_t:=\rho B_t/\sqrt c+\sigma_\perp U_t .
\]
The independent extension preserves the compensator of $\J$. Since
$D^\perp$ is independent of the original filtration,
$\langle B,D^\perp\rangle=0$. Hence the stochastic integral defining
$U$ has $\langle B,U\rangle=0$ and $\langle U\rangle=A$. Using
$\langle B\rangle=cA$ from \eqref{eq:limit-brackets-compensator}
and \eqref{eq:cov}, we obtain
\[
 \left\langle W'\right\rangle=\frac{\rho^2}{c}\langle B\rangle
                    +\left(1-\rho^2\right)\langle U\rangle=A,\qquad
 \left\langle B,W'\right\rangle=\frac{\rho}{\sqrt c}\langle B\rangle
                     =\rho\sqrt c\,A.
\]
Thus, $(B,W')$ and $\J$ satisfy the bracket and compensator identities
in \eqref{eq:limit-brackets-compensator}. The compensated mark integrals
are purely discontinuous and hence strongly orthogonal to the continuous
martingales $B,W'$. The variance equation \eqref{eq:target} involves only
$V,B,\J$ and holds on the extension. By the uniqueness of the joint law
in Theorem~\ref{thm:limit}, the log price in \eqref{eq:tprice} has the same
law as
\[
 X'_t=g_t+\sigma_\perp U_t,\qquad
 g_t=\rho B_t/\sqrt c-\Lambda J_t-\kappa A_t .
\]

Condition on $\mathcal H=\sigma(V,B,\J)$ over $[0,T]$.
The integrand of $U$ is then fixed, so $U$ has conditionally
independent Gaussian increments with variance $A_t-A_s$ on $[s,t]$.
For deterministic $0<r<T$, write
\[
 \inf_{r\le t\le T} X'_t
   =\sigma_\perp U_r+
     \inf_{r\le t\le T}\left\{g_t+\sigma_\perp(U_t-U_r)\right\}.
\]
Given $\mathcal H$, the two summands are independent. On $\{A_r>0\}$
the first has a Gaussian density with variance $\sigma_\perp^2 A_r>0$.
Consequently, for each $x\in\R$,
\begin{equation}\label{eq:barrier-tail-min}
 \Prob\!\left(A_r>0,\ \inf_{r\le t\le T} X'_t=x\right)=0 .
\end{equation}

Let $\tau:=\inf\{t\in[0,T]:A_t>0\}\wedge T$, with the convention $\inf\varnothing=\infty$.
Continuity of $A$ gives $A_\tau=0$. The stopped martingales $B^\tau$
and $U^\tau$ have zero quadratic variation, and the compensator gives
$\E[C_\tau]=m_0\E[A_\tau]=0$. Thus $X'_t=0$ up to and including $\tau$.
On $\{\inf_{0\le t\le T}X'_t=x\}$ for $x<0$, we have $\tau<T$.
By right continuity at $\tau$, there is a rational $r\in(\tau,T)$
with $\inf_{0\le t\le r}X'_t>x$. Then $A_r>0$ and
$\inf_{r\le t\le T}X'_t=x$.
Taking the countable union in \eqref{eq:barrier-tail-min} proves the assertion
for $x=\log(L/S_0)<0$.
\end{proof}


\subsection{Proof of Corollary~\ref{cor:path-prices}}\label{app:path-prices-proof}

\begin{proof}
Suppose $x_j\to x$ in $J_1$ on $[0,T]$. Choose increasing time
changes $\lambda_j$ onto $[0,T]$ with
$\|\lambda_j-\mathrm{id}\|_\infty+\|x_j\circ\lambda_j-x\|_\infty\to0$.
The paths $x_j$ are uniformly bounded, so for $s_j=S_0\e^{x_j}$ and
$s=S_0\e^x$ we have $\|s_j\circ\lambda_j-s\|_\infty\to0$. In particular,
\[
 \left|s_j(T)-s(T)\right|\ \vee\
 \left|\underline{s_j}_T-\underline s_T\right|
 \le \left\|s_j\circ\lambda_j-s\right\|_\infty .
\]
At every continuity point $t$ of $s$, $\lambda_j^{-1}(t)\to t$
gives $s_j(t)\to s(t)$. A c\`adl\`ag path has at most countably many
discontinuities, so the uniform bound on $s_j$ and the dominated convergence
theorem give $\overline{s_j}_T\to\overline s_T$. Applying the continuous
mapping theorem to the log-price coordinate of Theorem~\ref{thm:limit} yields
\[
 \left(S_T^h,\overline S^h_T,\underline S^h_T\right)
 \ \Longrightarrow\left(S_T,\overline S_T,\underline S_T\right).
\]
The arithmetic Asian put option payoff is a continuous function of the average.
For the barrier put option the discontinuity set is contained in
$\{\underline S_T=L\}$, which has probability zero by
Lemma~\ref{lem:barrier-atom}. Both put option payoffs lie in $[0,K]$,
so convergence in distribution gives convergence of their expectations.

For the Asian call option, the mean stock-price identity \eqref{eq:stock-means},
the grid average in \eqref{eq:grid-path-payoffs} and Tonelli's theorem yield
\[
 \E\left[\overline S^h_T\right]=\frac1N\sum_{n=0}^{N-1}\E\left[S^h_{t_n}\right]
 =S_0=\frac1T\int_0^T\E[S_t]\dd t=\E\left[\overline S_T\right].
\]
Apply the payoff identity $(x-K)^+-(K-x)^+=x-K$ to the grid average
$x=\overline S^h_T$ and the continuous-time average $x=\overline S_T$.
Taking expectations and using $\E[\overline S^h_T]=\E[\overline S_T]=S_0$ gives
\begin{align*}
 \E\left[\left(\overline S^h_T-K\right)^+\right]&=\E\left[\left(K-\overline S^h_T\right)^+\right]+S_0-K,\\
 \E\left[\left(\overline S_T-K\right)^+\right]&=\E\left[\left(K-\overline S_T\right)^+\right]+S_0-K .
\end{align*}
The convergence of the Asian put option prices therefore gives
\eqref{eq:path-price-convergence} for the Asian call option.
\end{proof}

\bibliographystyle{plainnat}
\bibliography{bibtex}

@article{hawkes,
  author = {Hawkes, Alan G.},
  title = {Spectra of some self-exciting and mutually exciting point processes},
  journal = {Biometrika}, year = {1971}, volume = {58}, pages = {83--90}
}

@article{hawkes-finance,
  author = {Hawkes, Alan G.},
  title = {{Hawkes} processes and their applications to finance: a review},
  journal = {Quantitative Finance}, year = {2018}, volume = {18}, number = {2},
  pages = {193--198}, doi = {10.1080/14697688.2017.1403131}
}

@article{bacry-microstructure,
  author = {Bacry, Emmanuel and Delattre, Sylvain and Hoffmann, Marc and Muzy, Jean-Fran\c{c}ois},
  title = {Modelling microstructure noise with mutually exciting point processes},
  journal = {Quantitative Finance}, year = {2013}, volume = {13}, number = {1},
  pages = {65--77}, doi = {10.1080/14697688.2011.647054}
}

@article{morariu-pakkanen,
  author = {Morariu-Patrichi, Maxime and Pakkanen, Mikko S.},
  title = {State-dependent {Hawkes} processes and their application to limit order book modelling},
  journal = {Quantitative Finance}, year = {2022}, volume = {22}, number = {3},
  pages = {563--583}, doi = {10.1080/14697688.2021.1983199}
}

@article{clusters,
  author = {Hawkes, Alan G. and Oakes, David},
  title = {A cluster process representation of a self-exciting process},
  journal = {Journal of Applied Probability}, year = {1974}, volume = {11}, pages = {493--503}
}

@article{inar,
  author = {Alzaid, A. A. and Al-Osh, M. A.},
  title = {An integer-valued {$p$}th-order autoregressive structure ({INAR}($p$)) process},
  journal = {Journal of Applied Probability}, year = {1990}, volume = {27}, pages = {314--324}
}

@article{kirchner,
  author = {Kirchner, Matthias},
  title = {Hawkes and {INAR}($\infty$) processes},
  journal = {Stochastic Processes and their Applications}, year = {2016}, volume = {126},
  pages = {2494--2525}, doi = {10.1016/j.spa.2016.02.008}
}

@article{wangmarked,
  author = {Wang, Haixu}, title = {Limit theorems for a discrete-time marked {Hawkes} process},
  journal = {Statistics \& Probability Letters}, volume = {184}, year = {2022},
  pages = {109368}, doi = {10.1016/j.spl.2022.109368}
}

@article{hxmarked,
  author = {Horst, Ulrich and Xu, Wei},
  title = {Functional limit theorems for marked {Hawkes} point measures},
  journal = {Stochastic Processes and their Applications}, year = {2021}, volume = {134},
  pages = {94--131}, doi = {10.1016/j.spa.2020.12.002}
}

@article{hxjumps,
  author = {Horst, Ulrich and Xu, Wei},
  title = {The microstructure of stochastic volatility models with self-exciting jump dynamics},
  journal = {Annals of Applied Probability}, year = {2022}, volume = {32}, number = {6},
  pages = {4568--4610}, doi = {10.1214/22-AAP1796}
}

@article{hxzrough,
  author = {Horst, Ulrich and Xu, Wei and Zhang, Rouyi},
  title = {Convergence of heavy-tailed {Hawkes} processes and the microstructure of rough volatility},
  journal = {Mathematical Finance}, year = {2026},
  doi = {10.1111/mafi.70064}, note = {Early View}
}

@article{hxzspikes,
  author = {Horst, Ulrich and Xu, Wei and Zhang, Rouyi},
  title = {Path-dependent fractional {Volterra} equations and the microstructure of rough volatility
           models driven by {Poisson} random measures},
  journal = {Finance and Stochastics}, year = {2026},
  note = {To appear. \href{https://arxiv.org/abs/2412.16436}{arXiv:2412.16436}}
}

@article{jr15,
  author = {Jaisson, Thibault and Rosenbaum, Mathieu}, title = {Limit theorems for nearly unstable {Hawkes} processes},
  journal = {Annals of Applied Probability}, year = {2015}, volume = {25}, pages = {600--631}
}

@article{jr16,
  author = {Jaisson, Thibault and Rosenbaum, Mathieu},
  title = {Rough fractional diffusions as scaling limits of nearly unstable heavy tailed {Hawkes} processes},
  journal = {Annals of Applied Probability}, year = {2016}, volume = {26}, pages = {2860--2882}
}

@article{rough,
  author = {Gatheral, Jim and Jaisson, Thibault and Rosenbaum, Mathieu}, title = {Volatility is rough},
  journal = {Quantitative Finance}, year = {2018}, volume = {18}, pages = {933--949}
}

@article{leverage,
  author = {El Euch, Omar and Fukasawa, Masaaki and Rosenbaum, Mathieu},
  title = {The microstructural foundations of leverage effect and rough volatility},
  journal = {Finance and Stochastics}, year = {2018}, volume = {22}, pages = {241--280}
}

@article{jrimpact,
  author = {Jusselin, Paul and Rosenbaum, Mathieu},
  title = {No-arbitrage implies power-law market impact and rough volatility},
  journal = {Mathematical Finance}, year = {2020}, volume = {30}, number = {4},
  pages = {1309--1336}, doi = {10.1111/mafi.12254}
}

@article{rh,
  author = {El Euch, Omar and Rosenbaum, Mathieu}, title = {The characteristic function of rough {Heston} models},
  journal = {Mathematical Finance}, year = {2019}, volume = {29}, pages = {3--38}, doi = {10.1111/mafi.12173}
}

@article{affinevolterra,
  author = {Abi Jaber, Eduardo and Larsson, Martin and Pulido, Sergio},
  title = {Affine {Volterra} processes},
  journal = {Annals of Applied Probability}, year = {2019}, volume = {29}, number = {5},
  pages = {3155--3200}, doi = {10.1214/19-AAP1477}
}

@article{abijaberL1,
  author = {Abi Jaber, Eduardo},
  title = {Weak existence and uniqueness for affine stochastic {Volterra} equations with {$L^1$}-kernels},
  journal = {Bernoulli}, year = {2021}, volume = {27}, number = {3},
  pages = {1583--1615}, doi = {10.3150/20-BEJ1284}
}

@article{cai,
  author = {Cai, Chunhao and He, Ping and Wang, Qinghua and Wang, Yingli},
  title = {Scaling limit of heavy-tailed nearly unstable cumulative {INAR}($\infty$) processes and rough fractional diffusions},
  journal = {Methodology and Computing in Applied Probability}, year = {2026}, volume = {28}, pages = {13}
}

@article{wcz,
  author = {Wang, Yingli and Cui, Zhenyu and Zhu, Lingjiong},
  title = {Rough {Heston} model as the scaling limit of bivariate cumulative heavy-tailed {INAR} processes:
           Weak-error bounds and option pricing},
  journal = {SIAM Journal on Financial Mathematics}, year = {2026},
  note = {To appear.}
}

@article{bps,
  author = {Bondi, Alessandro and Pulido, Sergio and Scotti, Simone},
  title = {The rough {Hawkes Heston} stochastic volatility model},
  journal = {Mathematical Finance}, year = {2024}, volume = {34}, pages = {1197--1241},
  doi = {10.1111/mafi.12432}
}

@incollection{askey-roy,
  author = {Askey, Richard A. and Roy, Ranjan},
  title = {Gamma Function},
  booktitle = {{NIST} Handbook of Mathematical Functions},
  editor = {Olver, Frank W. J. and Lozier, Daniel W. and Boisvert, Ronald F. and Clark, Charles W.},
  publisher = {Cambridge University Press}, year = {2010}, pages = {135--147},
  url = {https://dlmf.nist.gov/5.12.E1}
}

@misc{custers-asian,
  author = {Custers, Gijs and Karbach, Sven and Friesen, Martin},
  title = {Pricing and Hedging of Discretely Monitored {Asian} Options in the {Volterra--Heston} Model},
  year = {2026}, note = {\href{https://arxiv.org/abs/2609.07169}{arXiv:2609.07169}}
}

@mastersthesis{serva,
  author = {Serva, Giorgio Enrico Maria},
  title = {Numerical Methods and {SSR} for {Rough Hawkes Heston} Model},
  school = {Politecnico di Milano}, year = {2025},
  note = {\href{https://hdl.handle.net/10589/246994}{POLITesi repository}}
}

@misc{wwz,
  author = {Wang, Yingli and Wu, Yinhao and Zhu, Lingjiong},
  title = {Microstructural foundation for the rough {Hawkes--Heston} model},
  year = {2026}, note = {\href{https://arxiv.org/abs/2608.07709v1}{arXiv:2608.07709v1}}
}

@article{caravenna-doney,
  author = {Caravenna, Francesco and Doney, Ron},
  title = {Local large deviations and the strong renewal theorem},
  journal = {Electronic Journal of Probability}, year = {2019}, volume = {24},
  pages = {1--48}, note = {Paper No. 72}, doi = {10.1214/19-EJP319}
}

@book{billingsley,
  author = {Billingsley, Patrick}, title = {Convergence of Probability Measures}, edition = {2nd},
  year = {1999}, publisher = {Wiley}
}

@article{KarabashZhu,
  author = {Karabash, Dmytro and Zhu, Lingjiong},
  title = {Limit theorems for marked {H}awkes processes with application to a risk model},
  journal = {Stochastic Models}, 
  year = {2015}, 
  volume = {31}, 
  number = {3},
  pages = {433--451}
}

\end{document}